\documentclass[9pt,a4paper,twocolumn,twoside]{tau-class/tau}
\usepackage[english]{babel}
\usepackage{caption}
\usepackage{bm}
\usepackage{amsmath}
\usepackage{amssymb}
\usepackage{svg}
\usepackage{algorithm}
\usepackage{algpseudocode}
\usepackage{cancel}
\usepackage{booktabs}
\usepackage{multirow}
\usepackage{tabularx}
\usepackage{siunitx}
\usepackage{array}
\usepackage{braket}
\usepackage{makecell}
\usepackage{appendix}

\newcolumntype{C}{>{\centering\arraybackslash}X}
\newcolumntype{L}{>{\raggedright\arraybackslash}X}

\usepackage[normalem]{ulem}
\algrenewcommand\alglinenumber[1]{}

\algrenewcommand\algorithmicrequire{\textbf{Require:}}
\algrenewcommand\algorithmicensure{\textbf{Ensure:}}
\algrenewcommand\algorithmicforall{\textbf{for all}}
\algrenewcommand\algorithmicif{\textbf{if}}
\algrenewcommand\algorithmicthen{\textbf{then}}
\algrenewcommand\algorithmicelse{\textbf{else}}
\algrenewcommand\algorithmicend{\textbf{end}}
\algrenewcommand\algorithmicfor{\textbf{for}}
\algrenewcommand\algorithmicwhile{\textbf{while}}
\algrenewcommand\algorithmicreturn{\textbf{return}}

\journalname{arXiv Preprint}
\title{Physics-Informed Neural Networks for Maximizing Quantum Fisher Information in Time-Dependent Many-Body Systems}

\author[1,*]{Antonio Ferrer-Sánchez}
\author[1]{Yolanda Vives-Gilabert}
\author[2]{Yue Ban}
\author[2]{Xi Chen}
\author[1,3]{José D. Martín-Guerrero}

\affil[1]{Intelligent Data Analysis Laboratory (IDAL), Department of Electronic
Engineering, ETSE-UV, University of Valencia, Spain}
\affil[2]{Instituto de Ciencia de Materiales de Madrid ICMM-CSIC, 28049 Madrid, Spain}
\affil[3]{Valencian Graduate School and Research Network of Artificial Intelligence (ValgrAI), Spain}
\affil[4]{Kipu Quantum GmbH, Greifswalderstrasse 212, 10405 Berlin, Germany}

\professor{*e-mail: \href{mailto:antonio.ferrer-sanchez@uv.es}{Antonio.Ferrer-Sanchez@uv.es}}

\institution{Department of Electronic Engineering, ETSE-UV, University of Valencia.}
\footinfo{arXiv Preprint}
\theday{December 09, 2025}
\leadauthor{Antonio Ferrer-Sanchez et al.}

\begin{abstract}    
Quantum Fisher Information (QFI) sets the ultimate precision limit for parameter estimation and is therefore a central quantity in quantum metrology. In time-dependent many-body systems, however, maximizing QFI is a highly non-trivial task due to the combined effects of non-commutativity, control complexity, and the exponential growth of the Hilbert space. In this work, we present a physics-informed neural network (PINN) framework to address this problem through the learning of counter-diabatic quantum dynamics. Our approach combines a variational PINN formulation with a Magnus-expansion treatment of time-ordered evolution, enabling the adiabatic gauge potential and the scheduling function to be inferred directly from the underlying physics while enforcing the Euler-Lagrange structure of the protocol. The method is applied to several families of driven spin Hamiltonians, including nearest-neighbor, dipolar, and trapped-ion-inspired interactions, for systems of up to six qubits. The numerical results show that the proposed framework systematically improves over reference solutions based only on the Euler-Lagrange condition, yielding high normalized QFI together with favorable fidelity and extremal-balance metrics while preserving small phsical residuals. The analysis further shows that learning the scheduling function provides a clear performance advantage in most cases, and reveals non-trivial finite-size effects, with $q=3$ emerging as a particularly challenging regime. Although scalability remains limited by the exponential growth of the operator space and by automatic-differentiation costs, the results demonstrate that PINNs constitute a viable and physically grounded route for learning metrologically optimal control strategies in interacting quantum systems.
\end{abstract}


\begin{document}
		
    \maketitle 
    \thispagestyle{firststyle} 
    \tauabstract 
    
\section{Introduction}
\label{sec:Introduction}
\counterwithin*{equation}{section}
\renewcommand{\theequation}{1.\arabic{equation}}

Quantum Fisher Information (QFI), as a cornerstone of quantum metrology, provides the ultimate bound on the precision of parameter estimation through the quantum Cramér–Rao inequality~\cite{Braunstein-1994,Paris-2009}. In general, QFI quantifies how sensitively a quantum state responds to infinitesimal variations of a physical parameter, such as a magnetic field, coupling constant, or driving frequency. Maximization of QFI defines the upper limit of precision attainable in parameter estimation that, through optimization of QFI, effectively identifying the most sensitive quantum states and dynamical protocols capable of quantum-enhanced measurements towards the Heisenberg limit  \cite{Giovannetti-2006,Pezze-2009}.

Compared to the tremendous research devoted to the focus on time-independent Hamiltonian, a particularly rich setting, promising to broaden the application of quantum metrology into more complicated scenarios, arises when the Hamiltonian is explicitly time-dependent and contains a control or estimation parameter, i.e., $\pmb{\mathcal{H}}_{g}(t)$, where $g$ represents a measurable parameter. In such cases, both the eigenstructure and the dynamics of the system depend on the parameter of interest, and the time evolution itself acts as a metrological resource \cite{Jarzyna_2015}. The generator of parameter sensitivity, given by the derivative $\partial_{g}\pmb{\mathcal{H}}_{g}(t)$, directly governs the achievable QFI through the accumulated non-commutativity of $\pmb{\mathcal{H}}_{g}(t)$ at different times \cite{Kolodrubetz-2017,Deffner-2020}. Consequently, understanding how to design or modulate time-dependent parameters, such as driving amplitudes, frequencies, or interaction strengths, is central to achieving optimal precision in realistic quantum sensors. In particular, Pang and Jordan demonstrated that in single-qubit systems, optimal time-dependent control of the Hamiltonian can saturate the fundamental quantum limit of precision, establishing a general framework to maximize QFI in dynamically driven systems~\cite{Pang_2017}.
 
In the context of many-body quantum systems, this maximization problem becomes deeply non-trivial, since the intricate structure of entanglement, correlations, and criticality strongly affects how information about a parameter is encoded and redistributed across subsystems \cite{Zanardi-2008,Hyllus-2012}. For example, near quantum phase transitions, the QFI can exhibit scaling behaviors that reveal universal properties of critical dynamics \cite{Yin-2019,Hauke-2016,Rams-2018}. However, for general interacting systems, analytical evaluation and optimization of QFI are extremely challenging due to the exponential complexity of the Hilbert space.

In parallel, machine learning has become an effective and increasingly
popular tool for addressing complex problems in quantum metrology. 
Neural networks (NNs) allow for the use of quantum sensors even in regimes where the output response becomes complex and even in the presence of large shot noise, thus extending its working regime~\cite{Ban-2021,}. Employing a NN that replaces the computation of a physical observable offers a drastic reduction in the number of sample statistics~\cite{Torlai_2020}. Physics-Informed Neural Networks (PINNs)~\cite{seminal_pinns}, have emerged as powerful tools to solve time-dependent quantum dynamics subject to physical constraints. By embedding the Schrödinger equation and QFI maximization objective directly into a neural-network-based variational framework, one can efficiently explore the space of time-dependent controls even in many-body settings.

In this work, we address the maximization of QFI in time-dependent many-body Hamiltonians by explicitly incorporating the parametric dependence of the evolution of the system. Using a combination of the Magnus expansion for accurate unitary propagation and a PINN-based optimization framework, we show how temporal control and parameter modulation can be jointly designed to reach or approach the theoretical upper bounds of QFI. This approach not only provides a suitable computational route for QFI optimization in interacting quantum systems but also offers physical insight into how time-dependent Hamiltonians encode and amplify metrological sensitivity. Unlike standard optimal control approaches that typically optimize fidelity or energy functionals, our framework directly targets metrological performance by maximizing QFI. We use a PINN because it allows us to enforce the Schrödinger dynamics, the Euler-Lagrange constraint, and the QFI objective within a single continuous-time framework. By contrast, Gradient Ascent Pulse Engineering (GRAPE)~\cite{Khaneja_2005}, basis-expansion optimal-control methods such as CRAB (Chopped Random Basis)~\cite{Caneva_2011}, and reinforcement learning (RL) are more generic control optimizers and do not incorporate this physics-informed structure as directly.

This paper is structured as follows. First, we present the theoretical background and the counter-diabatic metrological framework. Next, we describe the proposed PINN-based methodology, including the trainable scheduling function, the Mangus treatment of time-ordered evolution, and the physics-informed loss. Afterwards, we report the numerical results for different Hamiltonian dynamics and system sizes, together with robustness and scalability analyses. Finally, we summarize the main conclusions and outline future directions.

\subsection{Theoretical background}
\label{subsec:background}

\textbf{From classical to quantum Fisher information.} From the perspective of classical metrology, the classical Fisher information (CFI) provides a quantitative measure of how much information a given random variable, denoted by $U$, carries about an unknown parameter $g$ that influences its probability distribution and is characterized by the standard deviation $\Delta g=\sqrt{\braket{g^{2}}-\braket{g}^{2}}$. In the context of quantum metrology, the measurement process can be formulated by treating $u$ as a classical measurement outcome and introducing a family of Hermitian operators $\{\hat{E}(u)\}$, commonly referred to as the positive operator-valued measure (POVM) associated with the outcome $u$. The complete measurement is a collection $\{\hat{E}(u)\}_{u}$ that satisfies,

\begin{equation}
\hat{E}(u)\geq 0,\qquad\sum_{u}\hat{E}(u)=\mathbb{I},
\label{eq:E_measurements}
\end{equation}

\noindent where $\hat{E}(u)$ encodes the quantum effect corresponding to ``the apparatus reports outcome $u$''. Having that $P(u|g)$ refers to the probability of measuring $u$ given parameter $g$ with $\sum_{u}P(u|g)=1$, the CFI can be defined as,

\begin{equation}
F_{C}\left(g;\hat{E}\right)=\sum_{u}\frac{1}{P(u|g)}\left(\frac{\partial P(u|g)}{\partial g}\right)^{2},\qquad P(u|g)=\mathrm{Tr}\left[\rho(g)\:\hat{E}(u)\right].
\label{eq:CFI}
\end{equation}

Here, $\rho(g)$ should be understood as the density matrix that encodes the information about the possible outcomes of measurements performed on the physical system under consideration; consequently, the trace-based equality is interpreted most naturally from a quantum-mechanical standpoint.

Regarding the precision with which the parameter $g$ can be estimated, its standard deviation is bounded by~\eqref{eq:Cramer_Rao}, known as the Cramér-Rao bound (CRB), where $\nu$ denotes the number of independent and identically distributed (i.i.d.) samples (trials). Thus, $F_{C}\left(g;\hat{E}\right)$ (often written simply as $F_{C}(g)$) quantifies the best achievable precision: larger Fisher information corresponds to smaller achievable variances. This quantity depends on the specific experiment---and on the measurement itself---since different conditional probabilities $P(u|g)$ (i.e., different physical setups) yield different values of $F_{C}(g)$.

\begin{equation}
\Delta g=\sqrt{\mathrm{Var}(\hat{g})}\geq\frac{1}{\sqrt{\nu\:F_{C}(g)}}.
\label{eq:Cramer_Rao}
\end{equation}

The quantum version of the Fisher information arises naturally when one interprets it as an upper bound obtained by maximizing $F_{C}(g)$ over all possible POVMs $\hat{E}(u)$ for the quantum systems under consideration. Consequently, the quantum counterpart of this limit---the quantum Cramér-Rao bound (QCRB)---can be defined as in~\eqref{eq:Cramer_Rao} by simply replacing the Fisher information with the quantum formulation in~\eqref{eq:QFI_1}.

\begin{equation}
F_{\mathrm{Q}}(g)=\max_{\hat{E}}\:F_{\mathrm{Q}}\left(g;\hat{E}\right).
\label{eq:QFI_1}
\end{equation}

\textbf{From general QFI to the pure-state formula.} We focus on a one-parameter family of pure states $\ket{\psi(g)}$, with $g\in\mathbb{R}$, normalized so that $\braket{\psi(g)|\psi(g)}=1$ for all $g$. In the most general (mixed-state) setting, the quantum Fisher information can be expressed in terms of the symmetric logarithmic derivative (SLD) $J_{g}$, defined by the Lyapunov equation~\eqref{eq:SLD_Lyapunov}, while the QFI itself can be written in a general form as~\eqref{eq:QFI_SLD}.

\begin{equation}
\partial_{g}\rho(g)=\frac{1}{2}\left(J_{g}\:\rho(g)+\rho(g)\:J_{g}\right).
\label{eq:SLD_Lyapunov}
\end{equation}

\begin{equation}
F_{\mathrm{Q}}(g)=\mathrm{Tr}\left[\rho(g)\:J_{g}^{2}\right].
\label{eq:QFI_SLD}
\end{equation}

This representation makes clear that $F_{\mathrm{Q}}(g)$ is an intrinsic property of the parametric family $\rho(g)$ (i.e., it is independent of any particular POVM), and it coincides with the optimal classical Fisher information over all measurements.

A particularly transparent expression is obtained for pure states, $\rho(g)=\ket{\psi(g)}\bra{\psi(g)}$. In this case, one can see how QFI depends only on the component of $\ket{\partial_{g}\psi(g)}$ orthogonal to $\ket{\psi(g)}$ since parallel changes to $\ket{\psi(g)}$ correspond to an unobservable global phase (a gauge degree of freedom). In this scenario, the QFI with respect to the parameter $g$, is reduced to

\begin{equation}
F_{\mathrm{Q}}(g)=4\left[\braket{\partial_{g}\psi(g)|\partial_{g}\psi(g)}-\left|\braket{\psi(g)|\partial_{g}\psi(g)}\right|^{2}\right].
\label{eq:QFI_pure}
\end{equation}

The first term in~\eqref{eq:QFI_pure} quantifies how the state changes relative to $g$, while the second term subtracts the phase-like component that does not change the physical ray in Hilbert space.

\textbf{Interrogation as unitary dynamics and the generator of parameter translations.} In quantum metrology, the parameter $g$ is typically \emph{encoded dynamically} during an interrogation stage: an initial state $\ket{\Psi_{\mathrm{in}}}:=\ket{\Psi_{g}(t=0)}$ evolves up to a final time $T$ under a (generally) time-dependent Hamiltonian operator, denoted as $\pmb{\mathcal{H}}_{g}(t)$, that depends on the parameter. The general dynamics are written in~\eqref{eq:U_def}, where the unitary time-evolution operator is defined through the time-ordering operator, $\mathcal{T}$, which takes into account the non-commutative contributions of the Hamiltonian through different time steps~\cite{Blanes_2009}.

\begin{equation}
\ket{\Psi_{g}(T)}=\pmb{\mathcal{U}}_{g}(T)\ket{\Psi_{\mathrm{in}}},\qquad\pmb{\mathcal{U}}_{g}(T)=\mathcal{T}\:\mathrm{exp}\left(-i\int_{0}^{T}\pmb{\mathcal{H}}_{g}(t)\:dt\right).
\label{eq:U_def}
\end{equation}

A central object in this setting is the Hermitian generator~\eqref{eq:h_def}, which captures how sensitively the \emph{entire evolution operation} depends on $g$ at the final time.

\begin{equation}
\pmb{h}_{g}(T)=i\:\pmb{\mathcal{U}}_{g}^{\dagger}(T)\:\left[\partial_{g}\pmb{\mathcal{U}}_{g}(T)\right].
\label{eq:h_def}
\end{equation}

For a pure initial probe state, the QFI admits a variance expression~\eqref{eq:QFI_variance} which can be straightforwardly derived from~\eqref{eq:QFI_pure}.

\begin{equation}
F_{\mathrm{Q}}(g;T)=4\left[\braket{\Psi_{\mathrm{in}}|\pmb{h}^{2}_{g}(T)|\Psi_{\mathrm{in}}}-\left|\braket{\Psi_{\mathrm{in}}|\pmb{h}_{g}(T)|\Psi_{\mathrm{in}}}\right|^{2}\right]=4\left(\Delta\pmb{h}_{g}(T)\right)_{\ket{\Psi_{\mathrm{in}}}}.
\label{eq:QFI_variance}
\end{equation}

This identity can be seen as the dynamical analogue of the familiar time-independent relation $F_{\mathrm{Q}}(g)=4\left(\Delta\pmb{\mathcal{H}}\right)^{2}$ for $\pmb{\mathcal{U}}_{g}=e^{-i\:g\pmb{\mathcal{H}}}$. For time-dependent Hamiltonians, $\pmb{h}_{g}(T)$ can be written in an integral form~\eqref{eq:h_integral} by breaking the unitary evolution $\pmb{\mathcal{U}}_{g}\left(0\to T\right)$ into products of small time intervals $\Delta t$ with $\Delta t\to 0$, thus exposing the role of the instantaneous operator $\partial_{g}\pmb{\mathcal{H}}_{g}(t)$, also known as the \emph{sensitivity operator} (see~\cite{Pang_2014}).

\begin{equation}
\pmb{h}_{g}(t)=\int_{0}^{T}\pmb{\mathcal{U}}_{g}^{\dagger}\left(0\to t\right)\:\partial_{g}\pmb{\mathcal{H}}_{g}(t)\:\pmb{\mathcal{U}}_{g}\left(0\to t\right)\:dt.
\label{eq:h_integral}
\end{equation}

Here, $\pmb{\mathcal{U}}_{g}\left(0\to t\right)$ denotes the temporal evolution from $0$ to a certain time step $t$, having $t\in[0,T]$. This representation is particularly useful for connecting metrological sensitivity directly to the time history of the \emph{parameter-derivative Hamiltonian} transported into the interaction scenario defined by the dynamics itself.

\textbf{Upper bound on QFI and extremal eigenspaces.} The relationship in~\eqref{eq:QFI_variance} implies that a large QFI requires a large separation between the eigenvalues of $\pmb{h}_{g}(T)$ that are populated by the probe state $\ket{\Psi_{\mathrm{in}}}$. Regarding time-dependent dynamics, it is possible to further bound the maximum achievable QFI by relating the spectrum of the said $\pmb{h}_{g}(T)$ to the instantaneous spectrum of the corresponding sensitivity operator, $\partial_{g}\pmb{\mathcal{H}}_{g}(t)$. Let $\lambda_{\mathrm{max}}(t)$ and $\lambda_{\mathrm{min}}(t)$ be the largest and smallest eigenvalues of $\partial_{g}\pmb{\mathcal{H}}_{g}(t)$. It is possible then to obtain the bound,

\begin{equation}
F_{\mathrm{Q}}(g;T)\leq\left[\int_{0}^{T}\left(\lambda_{\mathrm{max}}(t)-\lambda_{\mathrm{min}}(t)\right)\: dt\right]^{2}.
\label{eq:QFI_upper_bound}
\end{equation}

Hence, the absolute scale of metrological performance is defined by the \emph{time-integrated gap} between the extremal eigenvalues of the sensitivity operator. The bound in~\eqref{eq:QFI_upper_bound} is not only a formal conclusion: it indicates how a theoretical optimal protocol should be. If one can engineer the evolution such that, throughout the entire interval $t\in[0,T]$, the evolving components of $\ket{\Psi_{\mathrm{in}}}$ remain supported on the instantaneous eigenstates of $\partial_{g}\pmb{\mathcal{H}}_{g}(t)$, then the generator $\pmb{h}_{g}(T)$ at the end of the evolution will accumulate the largest possible spectral gap, causing the QFI to approach its maximal value. Let $\pmb{\mathcal{H}}_{\mathrm{tot}}(t)$ be the total time-dependent Hamiltonian governing the dynamics, and let $\ket{\phi_{\mathrm{min}}(t)}$ and $\ket{\phi_{\mathrm{max}}(t)}$ be the time-evolved extremal eigenstates of $\partial_{g}\pmb{\mathcal{H}}_{g}(t)$. Then, this idealized requirement can be interpreted as considering that the physical state of the system, $\ket{\Psi(t)}$, that undergoes the dynamics remains equal for the entire temporal evolution, to the superposition of the former, i.e.

\begin{equation}
\ket{\Psi(t)}=\frac{1}{\sqrt{2}}\left(\ket{\phi_{\mathrm{min}}(t)}+\ket{\phi_{\mathrm{max}}(t)}\right).
\label{eq:Psi_def}
\end{equation}

This condition~\eqref{eq:Psi_def} must be satisfied to guarantee the obtaining of the maximum QFI at the end of the evolution, $t=T$, where measurements take place. In this scenario, $\ket{\phi_{\mathrm{min}}(t)}=\pmb{\mathcal{U}}(0\to t)\ket{\phi_{\mathrm{min}}(0)}$ and $\ket{\phi_{\mathrm{max}}(t)}=\pmb{\mathcal{U}}(0\to t)\ket{\phi_{\mathrm{max}}(0)}$, where $\pmb{\mathcal{U}}(t)$ is understood as the time-evolution operator relying on $\pmb{\mathcal{H}}_{\mathrm{tot}}(t)$, which may (or may not) contain additional terms beyond $\pmb{\mathcal{H}}_{g}(t)$.

\textbf{Non-commutativity and counter-diabatic driving.} In generic many-body problems, two obstructions prevent the ideal protocol described above from being realized by simply evolving with $\pmb{\mathcal{H}}_{g}(t)$:

\begin{itemize}
    \item $\pmb{\mathcal{H}}_{g}(t)$ and $\partial_{g}\pmb{\mathcal{H}}_{g}(t)$ typically do not commute, so the eigenstates of the last are not preserved by natural dynamics.
    \item In general, $\pmb{\mathcal{H}}_{g}(t)$ at different times does not commute with itself, making time ordering essential and complicating attempts to keep the probe confined to special instantaneous subspaces.
\end{itemize}

As a result, even if the probe $\ket{\Psi_{\mathrm{in}}}$ is initialized using the extreme eigenstates of $\partial_{g}\pmb{\mathcal{H}}_{g}(0)$, the subsequent evolution over $\pmb{\mathcal{H}}_{g}(t)$ will generally induce leaks in other directions, degrading the accumulated sensitivity and preventing the QFI from achieving its maximum theoretical value. To address this drawback, we propose introducing a \emph{control Hamiltonian} operator, denoted $\pmb{\mathcal{H}}_{\mathrm{c}}(t)$, which is independent of the parameter of interest $g$, so that $\partial_{g}\pmb{\mathcal{H}}_{\mathrm{tot}}(t)=\partial_{g}\pmb{\mathcal{H}}_{g}(t)$, where $\pmb{\mathcal{H}}_{\mathrm{tot}}(t)=\pmb{\mathcal{H}}_{g}(t)+\pmb{\mathcal{H}}_{\mathrm{c}}(t)$ is the total Hamiltonian of the evolution. This is achieved by ensuring that the quantum state under consideration, $\ket{\Psi(t)}$, evolves in time according to the Schrödinger equation,

\begin{equation}
\pmb{\mathcal{H}}_{\mathrm{tot}}\ket{\Psi(t)}=i\partial_{t}\ket{\Psi(t)}.
\label{eq:Schrödinger}
\end{equation}

Within the shortcuts to adiabaticity (STA) philosophy~\cite{Odelin_2019}, the control is naturally implemented via a counter-diabatic (CD) term, written in terms of a scheduling function $\lambda(t)$ and an adiabatic gauge potential (AGP), denoted as $\pmb{\mathcal{A}}_{\lambda}(t)$. In general, this scheduling is chosen so that it reaches its maximum at intermediate times during the evolution while vanishing at the endpoints (initial and final times). Moreover, according to the general CD driving~\cite{Kolodrubetz-2017}, the AGP is an operator that does not induce non-zero transition amplitudes between instantaneous energy levels of $\pmb{\mathcal{H}}_{g}(t)$. Therefore, the total Hamiltonian governing the evolution is defined---and can be understood henceforth---as

\begin{equation}
\pmb{\mathcal{H}}_{\mathrm{tot}}(t)=\pmb{\mathcal{H}}_{g}(t)+\frac{\mathrm{d}\lambda}{\mathrm{d}t}\pmb{\mathcal{A}}_{\lambda}(t).
\label{eq:Htot_CD}
\end{equation}

With the dynamics specified in~\eqref{eq:Htot_CD}, the probe state at the end of the evolution, i.e. at $t=T$, can be written as follows, making a direct analogy with~\eqref{eq:U_def} but this time

With the dynamics defined in~\eqref{eq:Htot_CD}, the final state of the evolution, i.e. at $t=T$, can be defined as follows, by analogy with~\eqref{eq:U_def}. In this case, however, the time evolution is taken to be governed by the total Hamiltonian of the system, once the control term has been included.

\begin{equation}
\ket{\Psi(T)}=\pmb{\mathcal{U}}(T)\ket{\Psi_{\mathrm{in}}},\qquad\pmb{\mathcal{U}}(T)= \mathcal{T}\:\mathrm{exp}\left(-i\int_{0}^{T}\pmb{\mathcal{H}}_{\mathrm{tot}}(t)\:dt\right).
\label{eq:Psi_T}
\end{equation}

The metrological figure of merit we ultimately care about is the quantum Fisher information of this \emph{final} state with respect to the parameter $g$. Similarly, by analogy with~\eqref{eq:QFI_pure}, this magnitude can be written as follows, since $\ket{\Psi(T)}$ is a pure state.

\begin{equation}
F_{\mathrm{Q}}(g;T)=4\:\left[\braket{\partial_{g}\Psi(T)|\partial_{g}\Psi(T)}-\left|\braket{\Psi(T)|\partial_{g}\Psi(T)}\right|^{2}\right].
\label{eq:QFI_final}
\end{equation}

Within this framework, it is important to note that the adiabatic theorem~\cite{Born_1928,Avron_1988} ensures that the energy of a quantum system will remain close to that of its initial state, provided that the elapsed evolution time is sufficiently long for the transformation to proceed smoothly. On the other hand, a general time-dependent Hamiltonian---such as $\pmb{\mathcal{H}}_{g}(t)$---may induce changes in the quantum states it governs throughout the temporal evolution. Consequently, it is always possible to define an effective total Hamiltonian operator, as in~\eqref{eq:Htot_CD}, which is directly correlated with the previous and accurately reproduces the time progression of its instantaneous eigenstates. Therefore, the operator $\pmb{\mathcal{A}}_{\lambda}(t)$ can be understood as the set of terms responsible for capturing counter-diabatic effects, thereby enabling the acceleration of dynamical systems through the introduction of accessible degrees of freedom. Therefore, these terms can facilitate such a speed-up, achieving the same outcome as a fully adiabatic evolution while suppressing any undesired transitions between eigenstates~\cite{Odelin_2019}. As written in~\eqref{eq:H_CD}, the AGP is weighted by the time derivative of the scheduling function $\lambda(t)$, which can be interpreted as a measure of the overall transition rate.

\begin{equation}
\pmb{\mathcal{H}}_{\mathrm{CD}}(t)\equiv\frac{\mathrm{d}\lambda}{\mathrm{d} t}\pmb{\mathcal{A}}_{\lambda}(t).
\label{eq:H_CD}
\end{equation}

Although in the literature where this function is introduced within the CD protocols, it may be regarded as a collection of multiple components, in line with recent developments in the field, we are considering a \emph{scalar}-time parameterization whose purpose is to describe how the system is driven from a given initial Hamiltonian to a final (target) one. Using these lines, interpreting $\lambda(t)$ as an interpolation, it should in general satisfy: (i) $\lambda(t=0)=0$; (ii) $\lambda(t=T)=1$, where $T$ denotes the total evolution time; (iii) $\partial_{t}\lambda(t=0)=\partial_{t}\lambda(t=T)=0$; and (iv) $\lambda\in[0,1]$. It is straightforward to understand this interpolation behavior by checking how the control Hamiltonian is smoothly driven from an initial to a final operator~\eqref{eq:H_g}; generally, these can also be time-dependent.

\begin{equation}
\pmb{\mathcal{H}}_{g}(t)=\left(1-\lambda(t)\right)\:\pmb{\mathcal{H}}_{\mathrm{initial}}(t)+\lambda(t)\:\pmb{\mathcal{H}}_{\mathrm{final}}(t).
\label{eq:H_g}
\end{equation}

With regard to the AGP, and in accordance with the literature, the quantum system is assumed to evolve under a minimum-action principle, namely by minimizing the action function in~\eqref{eq:action} for the Hilbert-Schmidt operator $\pmb{\mathcal{G}}_{\lambda}(t)$. For a detailed derivation, the interested reader is referred to the methods section of~\cite{Dries_2017}.

\begin{equation}
\mathcal{S}=\mathrm{Tr}\:\left[\pmb{\mathcal{G}}^{2}_{\lambda}\right],\qquad\pmb{\mathcal{G}}_{\lambda}\left(\pmb{\mathcal{A}}_{\lambda}\right)=\partial_{\lambda}\pmb{\mathcal{H}}_{g}+i\left[\pmb{\mathcal{A}}_{\lambda},\pmb{\mathcal{H}}_{g}\right].
\label{eq:action}
\end{equation}

Consequently, the minimization of this action can be shown to yield the equation in~\eqref{eq:Euler-Lagrange}. This expression may therefore be understood as the Euler-Lagrange equation of the system, which is of central importance: only those operators $\pmb{\mathcal{A}}_{\lambda}(t)$ that satisfy it can be regarded as physically plausible solutions for the full evolution.

\begin{equation}
\left[i\partial_{\lambda}\pmb{\mathcal{H}}_{g}(t)-\left[\pmb{\mathcal{A}}_{\lambda}(t),\pmb{\mathcal{H}}_{g}(t)\right],\pmb{\mathcal{H}}_{g}(t)\right]=0.
\label{eq:Euler-Lagrange}
\end{equation}

\subsection{Deep-learning-based methodologies}
\label{subsec:DL_methods}

In many areas of physics, systems of partial differential equations (PDEs) are solved using high-accuracy finite-difference methods (FDM), which typically proceed sequentially by iterating over the independent variables---such as time and/or space---depending on the specific problem. Finite-difference concepts extend naturally to the numerical simulation of quantum dynamics because both the Schrödinger equation (for closed systems) and master equations (for open systems) define time-dependent differential evolution laws. One typically introduces a temporal discretization and approximates the continuous generator by a sequence of short-time propagators so that the state vector or density operator is advanced iteratively along the grid. In this context, frameworks such as QuTiP~\cite{Qutip_2012} provide dedicated solvers whose internal workflows rely on time discretization and numerical integration of ordinary differential equations (ODEs) to evaluate time-dependent Hamiltonian operators and propagating quantum states. However, other options such as PennyLane~\cite{pennylane_2022} implement analogous dynamics through differentiable simulation and circuit-based constructions of time-ordered evolution, where finite-difference estimates of derivatives may be employed.

In this context, deep-learning (DL) models such as Physics-Informed Neural Networks (PINNs) have been shown in recent literature~\cite{seminal_pinns} to provide a credible alternative to classical finite-difference solvers. These approaches aim to solve differential-equation systems by training a neural architecture constrained by the governing physical laws---often referred to as \emph{physical inductive biases}---thereby steering the solution using prior theoretical insight into the problem. In this regard, the universal approximation theorem~\cite{NN_theorem} ensures that neural networks (NNs) can represent solutions to broad classes of continuous PDEs, although practical performance depends on factors such as network architecture and capacity, as well as formulation of a robust loss function, among others. From a general perspective, a PINN is a neural model defined by a set of trainable parameters, denoted by $\Theta$, comprising the weights and biases to be optimized, i.e. $\Theta=\{\pmb{W}_{k},\pmb{b}_{k}\}_{1\leq k\leq\mathrm{K}}$, where $\pmb{W}_{k}\in\mathbb{R}^{N_{k}\times N_{k-1}}$ represents the connection weights and $\pmb{b}_{k}\in\mathbb{R}^{N_{k}}$ the corresponding biases. This would correspond to the most \emph{vanilla} formulation of a PINN consisting of $\mathrm{K}$ fully-connected stacked layers, which are supposed to be linear and implement the operation in~\eqref{eq:linear_operation}. Here, $\pmb{z}_{k}$ denotes the output of the $k$-th layer of the network, while $\{\sigma_{k}\}_{1\leq k\leq\mathrm{K}-1}$ denotes the set of non-linear activation functions that provide the model with expressive capacity, preventing it from reducing to a mere composition of linear maps. In general, these activations can vary across layers and can even be treated as trainable---either per layer or even per neuron~\cite{jagtap_2020}. They are typically applied to all layers except the final one, since the PINN output is usually interpreted as some physical quantity of interest and may need to satisfy problem-specific constraints that require an appropriate output mapping.

\begin{equation}
\pmb{z}_{k}=\pmb{W}_{k}\:\sigma_{k-1}\left(\pmb{z}_{k-1}\right)+\pmb{b}_{k}.
\label{eq:linear_operation}
\end{equation}

\begin{equation}
\mathcal{L}=\pmb{\omega}_{\mathrm{phys.}}\mathcal{L}_{\mathrm{phys.}}+\sum_{i}\pmb{\omega}_{i}\mathcal{L}_{i}.
\label{eq:general_loss}
\end{equation}

\subsection{Review of recent literature}
\label{subsec:review}

Within the field of quantum metrology, understanding the behavior of the Fisher information associated with a given parameter has become a central focus of recent research. The QFI constitutes the geometric-statistical measure that sets the ultimate precision limit for estimating a parameter encoded in a quantum state~\cite{Braunstein-1994}. Through the QCRB, the variance of any unbiased estimator is lower bounded by the inverse of the QFI. Therefore, maximizing this quantity is directly equivalent to minimizing the corresponding estimation uncertainty. To maximize the Fisher information in a $q$-qubit system, the relevant resources comprise: (i) the structure of the initial probe state---including its entanglement properties, entanglement depth, and underlying symmetries; (ii) the available Hamiltonian control; (iii) the design of the optimal measurement strategy; and (iv) the implementation of noise-mitigation mechanisms.

Works such as~\cite{Pang_2014} establish a general theoretical framework for parameter estimation under time-dependent Hamiltonians, deriving the generator of infinitesimal parametric translations and an upper bound for the QFI. They show that, in a $q$-qubit system, suitable entanglement enables the optimal quadratic scaling, $F_{\mathrm{Q}}\propto q^{2}$, regardless of the nature of the parameter. Other studies, such as~\cite{Jing_2015}, analytically investigate the QFI for unitary dynamics generated by collective SU(2) operators, encompassing many-qubit scenarios such as spin precession in a magnetic field. They derive an explicit expression for the maximal achievable QFI as a function of the evolution time, showing that it can be decomposed into two additive contributions: a term that grows quadratically in time and an oscillatory component. These results pave the way for works such as~\cite{Pang_2017}, where the time dependence of the Hamiltonian governing the evolution of the system is actively modulated. The authors develop an adaptive control strategy that continuously adjusts the Hamiltonian to maximize the QFI with respect to a target parameter. They show that, in the absence of control, the fundamental time scaling in unitary dynamics is $F_{\mathrm{Q}}\sim T^{2}$, with $T$ the evolution time considered; however, an appropriately engineered time-dependent modulation enables a quartic scaling, $F_{\mathrm{Q}}\sim T^{4}$, establishing a benchmark result in the field. In more recent studies, the authors in~\cite{Garbe_2022} investigate critical quantum metrology in fully-connected spin models, analyzing the scaling of the QFI near quantum phase transitions, showing that Heisenberg scaling, $F_{\mathrm{Q}}\sim q^{2}$, can in principle be achieved at criticality; however, this requires sufficiently slow protocols in order to exploit critical slowing down, thereby clarifying the interplay between criticality, entanglement, and metrological gain.

From the perspective of modern machine-learning-inspired numerical techniques, works such as~\cite{Shu_2025} employ tensor-network methods to study QFI scaling directly in the thermodynamic limit. By developing an algorithm based on infinite matrix product states, the authors compute the QFI density for $q\to\infty$, overcoming the finite-size limitations of previous analyses and providing a numerical framework for determining asymptotic QFI scaling laws in many-qubit systems. Some other works in the field whose main aim is the usage of reinforcement learning (RL) techniques have demonstrated the potential of these methods for quantum metrology. The authors in~\cite{Xu_2019} showed that deep RL can efficiently discover time-dependent control strategies that improve QFI; with learned policies generalizing across different parameter regimes and thus avoiding repeated re-optimization. In~\cite{Schuff_2020} RL-based pulse optimization was applied to drive spin systems, achieving substantial improvements in QFI by adaptively mitigating decoherence and implementing noise-tailored spin-squeezing mechanisms. More recently,~\cite{Xiao_2022} employed deep RL with physics-inspired control ansätze to address time-dependent Hamiltonians, attaining state-of-the-art precision in both noiseless and noisy scenarios, and outperforming standard gradient-based methods. Lately, works such as~\cite{Laxmisha_2025} adopt a machine-learning approach to maximize the QFI in multi-qubit systems by optimizing the distribution of entanglement across quantum sensor networks. With their approach, also based on RL techniques, the authors design shallow quantum circuits that prepare highly entangled probe states with enhanced QFI, while incorporating realistic noise models and error mitigation (e.g., via Qiskit-based simulations). This strategy enables the automated discovery of hardware-efficient states that approach Heisenberg-limited sensitivity, highlighting the potential of machine learning for scalable quantum metrology. Together, these works establish RL as a powerful and flexible framework for optimizing control protocols in quantum sensing.

Regarding PINN-based methodologies, complementary works such as~\cite{Norambuena_2024,Sofiia_2025} have recently addressed quantum control problems within this framework. In these studies, the Lindblad master equation is incorporated directly into the loss function, enabling the learning of control fields that drive the system from a given initial state to a desired target state with high fidelity. Within the framework of STA, based on counter-diabatic driving, a substantial body of literature addresses the construction of CD terms through analytical and numerical techniques without involving deep learning (e.g.,~\cite{Sels_2017,Chandarana_2023,Barone_2024} and references therein). Building on this line of research,~\cite{Ferrer_2024} introduces a PINN-based approach to infer approximately optimal CD operators by enforcing the principle of least action and Hermiticity constraints within the loss function. The method achieves high-fidelity adiabatic state transfer in two-qubit systems, although scalability remains limited by current classical hardware resources. In this regard, the physical memory requirements associated with the classical simulation of an increasing number of qubits rapidly become impractical. In particular, for dense $n\times n$ matrices, the computation of the matrix exponential operation scales as $\mathcal{O}(n^{3})$. Therefore, for a system of $q$ qubits and $N_{t}$ time steps, evaluating and propagating the time evolution through a unitary operator such as that in~\eqref{eq:QFI_final} scales as $\mathcal{O}\left(N_{t}2^{3q}\right)$~\cite{Leung_2017}. In this way, the classical simulation of a $q$-qubit system exhibits a clear scalability limitation, since reproducing its time evolution under a given Hamiltonian on classical hardware entails several computational bottlenecks. These include: (i) the application of the time-evolution operator $\pmb{\mathcal{U}}(t)$; (ii) state propagation and circuit depth, as longer circuits require a correspondingly larger number of elementary operations; and (iii) the sparsity and locality structure of the Hamiltonian. Indeed, for operators with a favorable structure---i.e. with many vanishing or negligible contributions~\cite{Samuel_2022}---sparse-matrix techniques may be advantageous; however, this is not a general-purpose scenario.

\section{Methodology}
\label{sec:Methodology}
\counterwithin*{equation}{section}
\renewcommand{\theequation}{2.\arabic{equation}}

\begin{figure}[!h]
    \centering
    \includegraphics[width=1.0\columnwidth]{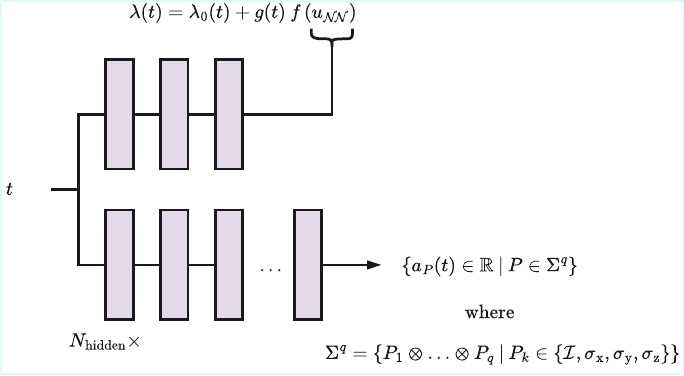}
    \caption{\justifying The proposed neural architecture for the PINN. The main set of stacked layers (below) processes the time coordinate input, $t$, and gives the set of $\{a_{P}(t)\}$ in which the $\pmb{\mathcal{A}}_{\lambda}(t)$ can be decomposed. On the other hand, the above path produces the $u_{\mathcal{NN}}$ output used to construct the scheduling $\lambda(t)$.}
    \label{fig:PINN_diagram}
\end{figure}

The main aim of our methodology is to construct a PINN capable of inferring the counter-diabatic contributions, i.e., the adiabatic gauge potential $\pmb{\mathcal{A}}_{\lambda}(t)$, as a function of time. The learned terms must be consistent with the physical constraints---respecting the Principle of Least Action, hermiticity of the operators, and the time-ordering of the evolution, among others---while driving the protocol to the maximization of the quantum Fisher information with respect to a certain control parameter, approaching as closely as possible the maximum achievable QFI~\eqref{eq:QFI_upper_bound}. Since the set of Pauli matrices $\Sigma:=\{\pmb{I},\sigma_{\mathrm{X}},\sigma_{\mathrm{Y}},\sigma_{\mathrm{Z}}\}\in\mathcal{M}_{2\times 2}
(\mathbb{C})$ constitutes an orthogonal basis for the Hilbert space of $2\times 2$ Hermitian matrices, any Hermitian operator acting on a $q$-qubit system can be written as a linear combination of tensor products of Pauli operators. Defining the set of Pauli strings~\eqref{eq:Pauli_strings}, one has the AGP expanded as~\eqref{eq:AGP_decomposition}.

\begin{equation}
\mathcal{P}_{q}:=\{P_{1}\otimes\ldots\otimes P_{q}\:|\:P_{k}\in\Sigma\}.
\label{eq:Pauli_strings}
\end{equation}

\begin{equation}
\pmb{\mathcal{A}}_{\lambda}(t)=\sum_{P\in\mathcal{P}_{q}}a_{P}(t)\:P,\qquad a_{P}(t)\in\mathbb{R}.
\label{eq:AGP_decomposition}
\end{equation}

In this representation, the set of coefficients $\{a_{P}(t)\}_{P\in\mathcal{P}_{q}}\in\mathbb{R}^{4^{q}}$ provides a symbolic parameterization of the AGP, avoiding the explicit reconstruction of matrix operators. Each coefficient is a time-dependent scalar function, and at most $4^{q}$ such functions are defined, corresponding to all possible Pauli-string combinations acting on a $q$-qubit system\footnote{Working with the complete six-qubit operator space is computationally prohibitive in general. Consequently, for the $q=6$ case, the AGP is not expanded over the full set of $4^{6}$ Pauli strings, but over a truncated retained operator basis up to the interactions between 4 qubits. In this way, we achieve a feasible computational implementation while still providing a tractable physically motivated operator.} Therefore, we introduce the architecture shown in figure~\ref{fig:PINN_diagram}, in which the coordinate time\footnote{It is important to note here that the time dimension can be normalized to the maximum time of evolution by defining $\tau=\frac{t}{T}\in[0,1]$. In that sense, the derivative with respect to $t$ shall be written as $\partial_{t}=\frac{1}{T}\partial_{\tau}$ in all calculations. However, for the sake of simplicity, we will consider $T=1$ (thus $t=\tau$) for all future simulations presented in the paper.} $t\in[0,1]$ is the input, and the set of $4^{q}$ coefficients in~\eqref{eq:AGP_decomposition} an output. It should be noted that, since the coefficients in the decomposition are real, the operator $\pmb{\mathcal{A}}_{\lambda}(t)$ is Hermitian by construction.

Here, the number of outputs that we would have in our model reads as~\eqref{eq:size_output}, where we have a fixed output, corresponding to $\lambda(t)$, where $q$ represents the number of qubits (system size), and where $k$ represents the maximum $k$-local interaction between qubits. It is therefore important to emphasize this point, since, in general, an operator such as the AGP, which can be decomposed in the Pauli-matrix algebra, may contain up to $4^{q}$ possible terms, corresponding to all possible combinations of interactions among the qubits. In practice, however, both in simulations and in real quantum hardware, one works only with a subset of these terms, typically by restricting the interactions to local contributions involving at most $k$ qubits. Consequently, when $k=q$, the equation below collapses to $N_{\mathrm{out}}=1+4^{q}$, which is preferable as long as the classical hardware that simulates the dynamics is able to handle the underlying processes.

\begin{equation}
N_{\mathrm{out}}(q,k)=1+\sum_{l=0}^{k}\frac{q!}{l!(q-l)!}3^{l}.
\label{eq:size_output}
\end{equation}

\subsection{Construction of the scheduling function}
\label{subsec:Scheduling_function}

As stated previously, the function $\lambda(t)$ drives the time evolution of the quantum system, and both $\lambda(t)$ and its derivative must satisfy specific boundary conditions at the beginning and at the end of the protocol. A common choice in the literature about counter-diabatic protocols is the functional form

\begin{equation}
\lambda(t)=\mathrm{sin}^{2}\left(\frac{\pi t}{2T}\right),
\label{eq:fixed_scheduling}
\end{equation}

\noindent where $T$ is the last step, understood as the \emph{measurement time}. This function poses a smooth and regular interpolation; its derivative attains a single maximum at the midpoint of the evolution, meaning that the counter-diabatic contribution to the total Hamiltonian is maximum at that step. Thus, the specific choice of this function is inherently somewhat arbitrary: it is typically fixed \emph{a priori} and may not be the profile that the network would give as ``optimal'' for the problem at hand. We therefore propose to provide the model with a trainable component, since the resulting $\lambda(t)$ satisfies the required boundary conditions and can still be considered as an interpolation function. This trainable schedule can be constructed as in~\eqref{eq:scheduling_function}. We start from a fixed (non-trainable) component, denoted by $\lambda_{0}(t)$, that satisfies the aforementioned conditions and will be referred to as the \emph{base schedule}. We then add an extra term---the \emph{correction envelope}---$c(t)=g(t)\:f\left(u_{\mathcal{NN}}\right)$, where both $g(t)$ and its derivative must vanish at the endpoints of the evolution; otherwise, the correction could contaminate $\lambda(t)$ at these critical points. A specific functional ansatz for $\lambda_{0}(t)$ and $g(t)$ is given in~\eqref{eq:scheduling_function__defs}.

\begin{equation}
\lambda(t)=\lambda_{0}(t)+c(t);\qquad c(t):=g(t)\:f\left(u_{\mathcal{NN}}\right).
\label{eq:scheduling_function}
\end{equation}

\begin{equation}
\lambda_{0}(t)=3t^{2}-2t^{3},\qquad g(t)=t^{2}(1-t)^{2}.
\label{eq:scheduling_function__defs}
\end{equation}

Regarding $f(\cdot)$, this is a function that acts directly on an output of the model, written here as $u_{\mathcal{NN}}$, and is not, in principle, bounded to any particular interval. However, $\lambda(t)$ must lie within $[0,1]$, so it is crucial to define $f(\cdot)$ in such a way that it respects this condition. To do so, we shall consider $|f(z)|\leq K$ for any real value $z$. Therefore, it suffices to choose a $K$ such that

\begin{equation}
K\:g(t)\leq\min\:\{\lambda_{0}(t),1-\lambda_{0}(t)|\}\qquad\forall t,
\label{eq:K_condition}
\end{equation}

i.e.

\begin{equation}
K \le \inf\limits_{t\in(0,1)} \frac{\min\:\{\lambda_0(t),\,1-\lambda_0(t)\}}{g(t)}.
\label{eq:lambda_infimum}
\end{equation}

where ``inf'' refers to the infimum. This demands $f(\cdot)$ respects

\begin{equation}
-\frac{3-2t}{(1-t)^{2}}\leq f\left(u_{\mathcal{NN}}(t)\right)\leq\frac{1+2t}{t^{2}}\qquad\forall\in(0,1).
\label{eq:restriction_f}
\end{equation}

If we split the entire time domain into two regions, $t\in\left[0,\frac{1}{2}\right)$ and $t\in\left[\frac{1}{2},1\right)$, it is straightforward to check that the infimum from both limits of $t\to 0^{+}$ and $t\to 1^{-}$ in each interval, respectively, gives the value of $K=3$ as the \emph{safe amplitude} for the particular definitions given in~\eqref{eq:scheduling_function__defs}. Therefore, the function must produce an absolute value no greater than $K$; here, \emph{tanh} seems to be the best fit, since $|\tanh|\leq 1$.

\begin{equation}
f(\cdot)=K\tanh(\cdot),\qquad K=3.
\label{eq:f}
\end{equation}

The construction in~\eqref{eq:f} ensures that the resulting $\lambda(t)$ satisfies all the requirements to be interpreted as a valid interpolation function. We note that alternative choices are certainly possible---both for $\lambda_{0}(t)$ and $g(t)$, as well as for the selection of $f(\cdot)$. Here, however, we delegate this flexibility to the PINN by letting it exploit the introduced degree of freedom to optimize the problem.

\subsection{Time-ordered evolution via the Magnus expansion}
\label{subsec:Magnus_expansion}

Broadly speaking, time evolution under a (generally) time-dependent Hamiltonian is implemented by propagating an initial state---$\ket{\Psi_{\mathrm{in}}}$---with the unitary operator $\pmb{\mathcal{U}}(t)$, as the one written in~\eqref{eq:Psi_T}. In practice, this operator is evaluated by discretizing time into $N_{t}$ steps of size $\Delta t$ and building an operator $\pmb{\mathcal{U}}\left(t\to t+\Delta t\right)$ which will drive the state from one time step to the next one. This involves, in general, an operator acting on a $2^{q}$-dimensional Hilbert space; the key scalability bottleneck is the storage of dense-matrix representations scale as $\mathcal{O}(4^{q})$, and matrix exponentials and multiplications scale as $\mathcal{O}(8^{q})$ per time step. Regarding the state propagation, it scales as $\mathcal{O}(4^{q})$ per step (matrix-vector), so runtime grows exponentially with $q$ and linearly with $N_{t}$, being $\mathcal{O}(N_{t}4^{q})$ for states and $\mathcal{O}(N_{t}8^{q})$ for dense unitary matrices. In addition, it is necessary to compute and backpropagate the gradients through all steps when training the model, adding additional multiplicative overhead. All these ingredients quickly lead to hardware overload, unless some previous algebra and approximations are considered.

To mitigate the prohibitive computational cost that scales poorly with the number of time steps, the Magnus expansion~\cite{Magnus_1954,Blanes_2009} has been widely adopted to approximate---yet provide sufficient accuracy in---the time evolution of quantum states. In this way, the operations required to apply the $N_{t}-1$ consecutive propagations that map the state from $t_{0}$, $\ket{\Psi(t=t_{0})}$, to $\ket{\Psi(t=T)}$, are drastically reduced by partitioning the time axis into a sequence of $n_{w}$ windows. In general, the $j$-th time step within the $w$-th window can be identified as~\eqref{eq:t_windows}. Here, the time evolves within $t\in[0,1]$ in all our numerical experiments, so $\Delta t=\frac{1}{N_{t}-1}$ can be seen as the discrete time step.

\begin{equation}
t^{(w)}_{j} \;=\; t_0 + \bigl(wm + j\bigr)\Delta t,\qquad 
w=0,\dots,n_{w}-1,\;\; j=0,\dots,m-1.
\label{eq:t_windows}
\end{equation}

In this context, the Magnus expansion allows us to express the time-evolution operator between successive time steps as a dense matrix exponential of an effective operator, denoted by $\pmb{\Omega}$. By partitioning the time axis into a set of windows, the computational cost of the propagation is reduced from sequentially evaluating all differential time evolutions---which scales linearly with $N_{t}$---to evaluating as many evolutions as the number of windows (i.e., the chosen subdivisions). This strategy removes direct access to the state $\ket{\Psi(t)}$ at every intermediate time instant; however, this limitation is typically not a matter of concern, since we are generally interested in the last point of the evolution, namely the state at $t=T$. Therefore, the $\pmb{\Omega}$ operator collects all contributions that arise from the fact that the Hamiltonian operator---$\pmb{\mathcal{H}}_{\mathrm{tot}}$ in our case---does not commute with itself at different time steps, i.e. $\left[\pmb{\mathcal{H}}_{\mathrm{tot}}(t_{1}),\pmb{\mathcal{H}}_{\mathrm{tot}}(t_{2})\right]\neq 0$. More generally, $\pmb{\Omega}$ can be written as a series expansion, where successive higher-order terms account for nested commutators of increasing order. For a given window $w$, we propagate the state $\ket{\Psi\left(t_{0}^{(w)}\right)}$ to the final time of that window by applying the operator in~\eqref{eq:Magnus_time_evolution}, while $\pmb{\Omega}^{(w)}$ is expanded as in~\eqref{eq:Magnus_series_discrete_truncated}. The three first orders are explicitly reported in equation~\eqref{eq:Magnus_terms} of ~\ref{sec:appendix1}.

\begin{equation}
\pmb{\mathcal{U}}\left(t_{0}^{(w)}\to t_{m-1}^{(w)}\right)=\mathrm{exp}\left(\pmb{\Omega}^{(w)}\right).
\label{eq:Magnus_time_evolution}
\end{equation}

\begin{equation}
\pmb{\Omega}^{(w)} \;\approx\; \sum_{n=1}^{p} \pmb{\Omega}^{(w)}_{n},
\qquad w=0,1,\dots,n_{w}-1.
\label{eq:Magnus_series_discrete_truncated}
\end{equation}

\begin{figure}[h]
    \centering
    \includegraphics[width=1.0\columnwidth]{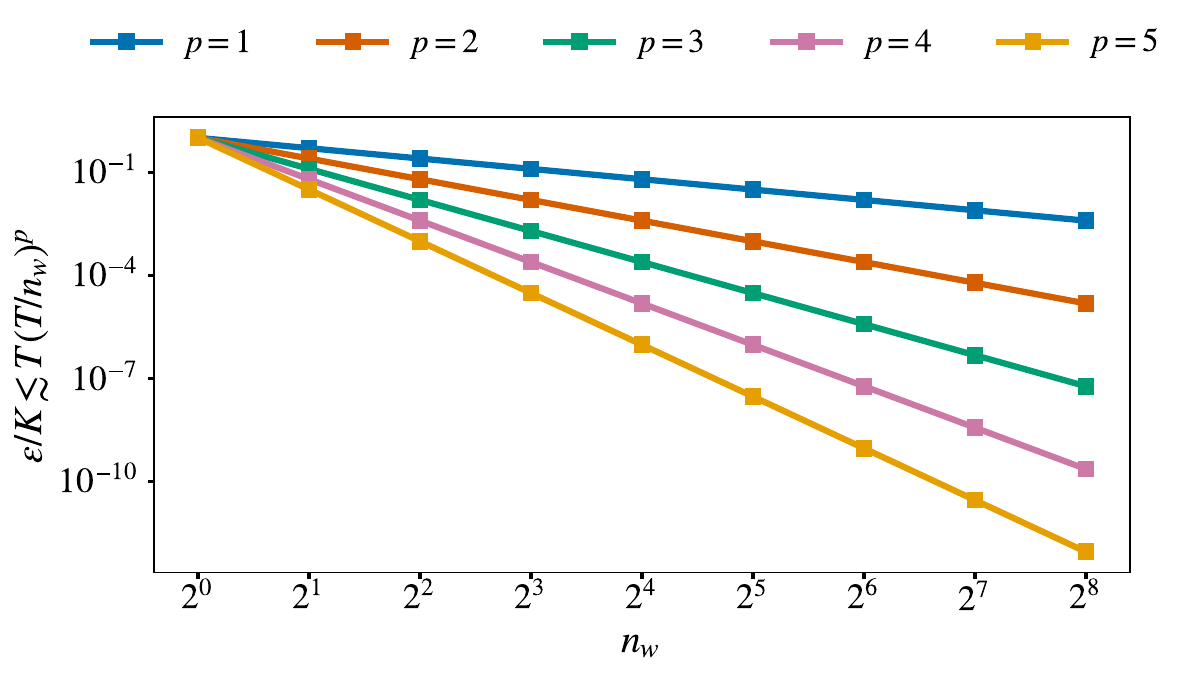}
    \caption{\justifying Global (fine-time) truncation error bound for a windowed Magnus propagator as a function of the number of time windows $n_{w}$ (with $T=1$); curves correspond to truncation order $p$ and illustrate the predicted scaling $\varepsilon/K\lesssim T(T/n_{w})^{p}$.}
    \label{fig:Magnus_error_methods}
\end{figure}

It is straightforward to see that as the expansion order increases, the approximation error decreases, albeit at the expense of a higher computational cost. As discussed in~\ref{sec:appendix1}, in our case this error goes as $\varepsilon\equiv\mathcal{O}\left(T(m\Delta t)^{p}\right)$, where $T$ is the total time evolution, $m$ the number of time steps within each window and $p$ corresponds to the order considered in the Magnus expansion. For instance, in a standard configuration, we typically set $N_{t}=2^{8}$, $T=1$, and $n_{w}=2^{4}$ (thus $m=2^{4}$). In this setting, the resulting error is approximately $\sim 2.24\times 10^{-4}$ for $p=3$, and $\sim 4\times 10^{-3}$ for $p=2$. This behavior is summarized in figure~\ref{fig:Magnus_error_methods}, where we plot the error $\varepsilon$, as derived in~\ref{sec:appendix1} and corresponding to equation~\eqref{eq:theo_error_Magnus_expansion}. One observes the scaling $\varepsilon\propto n_{w}^{-p}$, and, in particular, that for a sufficiently large number of windows, errors can be obtained comparable to those obtained at higher orders when using smaller values of $p$.

\subsection{Construction of the loss function}
\label{subsec:loss_function}

In constructing the total physics-informed loss that the PINN model minimizes during training, a first crucial component is the minimization of the physical action through the Euler-Lagrange equations introduced in~\eqref{eq:Euler-Lagrange}. This condition involves the AGP, $\pmb{\mathcal{A}}_{\lambda}(t)$; therefore, enforcing its minimization is essential, as it ensures that the $\pmb{\mathcal{A}}_{\lambda}(t)$ learned by the PINN is not an arbitrary solution, but rather one that adheres to the underlying physics of the CD protocol. Since the model outputs the coefficients of the basis elements used to decompose $\pmb{\mathcal{A}}_{\lambda}(t)$, it is natural to work with this expression directly in the Pauli representation, without involving dense matrix manipulations. Indeed, the relevant operations are commutators and the Pauli algebra provides well-defined commutation rules; as a result, this contribution to the loss can be evaluated with comparatively favorable scaling.

\begin{equation}
\mathcal{L}_{\mathrm{E-L}}(t)=\frac{1}{4^{q}}\left|\left[i\partial_{\lambda}\pmb{\mathcal{H}}_{g}(t)-\left[\pmb{\mathcal{A}}_{\lambda}(t),\pmb{\mathcal{H}}_{g}(t)\right],\pmb{\mathcal{H}}_{g}(t)\right]\right|^{2}.
\label{eq:Euler-Lagrange_loss_1}
\end{equation}

Here, it is important to note that since we work in coefficient space rather than in the dense-operator space, the expression in~\eqref{eq:Euler-Lagrange_loss_1} can be reduced to~\eqref{eq:Euler-Lagrange_loss_coeffs}, where the derivation of the coefficients $\{r_{k}(t)\}_{k=1}^{M}$ is detailed in~\ref{sec:appendix2}.

\begin{equation}
\mathcal{L}_{\mathrm{E-L}}(t)=\frac{1}{M}\sum_{k=1}^{M}\left|r_{k}(t)\right|^{2}.
\label{eq:Euler-Lagrange_loss_coeffs}
\end{equation}

In addition to the Euler-Lagrange equation, several other elements are going to be required, which we present below.

\textbf{(i) Final-values normalized quantum Fisher information.}

Since the QFI may vary across different system sizes and dynamical regimes, if our aim is to use this quantity not only as an evaluation metric but also as a training cost, it is convenient to have it normalized and lying within a well-defined and non-arbitrary interval. Following the definition of the maximum achievable QFI according to~\eqref{eq:QFI_upper_bound}, and noting that $\lambda_{\mathrm{min}}(t)$ and $\lambda_{\mathrm{max}}(t)$ are the extremal instantaneous eigenstates of $\partial_{g}\pmb{\mathcal{H}}_{g}(t)$, the normalized QFI at $t=T$ with respect to the parameter $g$ can be written as

\begin{equation}
\eta_{g}\equiv\frac{F_{Q}(g;T)}{F_{Q}^{\mathrm{max}}(g)}\in[0,1],\quad F_{Q}^{\mathrm{max}}(g)\equiv\left[\int_{0}^{T}\left(\lambda_{\mathrm{max}}(t)-\lambda_{\mathrm{min}}(t)\right)\:dt\right]^{2}.
\label{eq:final_QFI}
\end{equation}

This quantity is essential to enforce the maximization of this information, since $F_{Q}^{\mathrm{max}}(g)$ is calculated precisely for each particular case before training. Regarding $F_{Q}(g;T)$, it is computed during training according to~\eqref{eq:QFI_final}, where the subscript ``$g$'' in the quantum state was omitted for notational simplicity. Nevertheless, it should be kept in mind that $\ket{\Psi(t)}$ at any given time is associated with a specific value of $g$. Therefore, the variation of the state with respect to $g$, denoted $\ket{\partial_{g}\Psi_{g}(T)}$, can be computed using central numerical differences such as

\begin{equation}
\ket{\partial_{g}\Psi_{g}(T)}\approx\frac{\ket{\Psi_{g+\delta g}(T)}-\ket{\Psi_{g-\delta g}(T)}}{2\:\delta g},
\label{eq:dPsi_dg}
\end{equation}

\noindent where $\delta g$ must be small enough to provide an accurate approximation and the numerical calculation must be precise enough for this value to be considered reliable.

\textbf{(ii) Final-state fidelity.} According to the literature, the quantum Fisher information is expected to reach its maximum value when equality in~\eqref{eq:Psi_def} can be maintained throughout the entire evolution time $t$; that is, when the physical state of the system, denoted by $\ket{\Psi(t)}$, coincides with the normalized state obtained from the coherent superposition of the maximal eigenstates of the sensitivity operator with respect to the parameter $g$, namely $\partial_{g}\pmb{\mathcal{H}}_{g}(t)$, being these written as $\ket{\phi_{\mathrm{min}}(t)}$, $\ket{\phi_{\mathrm{max}}(t)}$. This target state at $t=T$ then reads

\begin{equation}
\ket{\Psi_{\mathrm{tar}}(T)}=\frac{1}{\sqrt{2}}\left(\ket{\phi_{\mathrm{min}}(T)}+\ket{\phi_{\mathrm{max}}(T)}\right).
\label{eq:state_target}
\end{equation}

However, from a practical standpoint, the intermediate time instants of the evolution where the CD protocol is applied are not accessible; therefore, the most pragmatic approach is to obtain the physical information at the measurement time. In our framework, both the initial state vector $\ket{\Psi_{\mathrm{in}}}$, and the target state are pure, since they are state vectors of a closed quantum system; accordingly, the fidelity at the final time can be computed as follows,

\begin{equation}
\mathcal{F}=\left|\braket{\Psi(T)|\Psi_{\mathrm{tar}}(T)}\right|^{2}\in[0,1].
\label{eq:final_state_fidelity_1}
\end{equation}

It is possible to define $p_{\mathrm{min}}$ and $p_{\mathrm{max}}$ at $t=T$ as the populations of these maximal states understood as the probabilities that $\ket{\Psi(T)}$ occupies each of the maximal eigenstates of $\partial_{g}\pmb{\mathcal{H}}_{g}(T)$~\eqref{eq:p_min__p_max}. Moreover, $\Delta\varphi$ can be described as the relative phase between the two extremal components of the final state. With these definitions, the fidelity can be rewritten as in~\eqref{eq:final_state_fidelity_2}.

\begin{equation}
p_{\mathrm{min}}\equiv\left|\braket{\Psi(T)|\phi_{\mathrm{min}}(T)}\right|^{2},\qquad p_{\mathrm{max}}\equiv\left|\braket{\Psi(T)|\phi_{\mathrm{max}}(T)}\right|^{2}.
\label{eq:p_min__p_max}
\end{equation}

\begin{equation}
\mathcal{F}=\frac{1}{2}\left(p_{\mathrm{min}}+p_{\mathrm{max}}+2\sqrt{p_{\mathrm{min}}p_{\mathrm{max}}}\mathrm{cos}(\Delta\varphi)\right).
\label{eq:final_state_fidelity_2}
\end{equation}

Regarding the relative phase, when $\Delta\varphi=0$, the two components are in phase, which is optimal for the target state; by contrast, when $\Delta\varphi=\pi$, the components are out of phase. Intermediate values lead to partially constructive or destructive interference. It will therefore be essential to keep this quantity under control in our methodology.

\textbf{(iii) Final-state extremal-balance score.} In addition to the fidelity, the following metric $\mathcal{B}$ is introduced to quantify the balance between the two extremal eigenstates of the sensitivity operator. This quantity~\eqref{eq:B} reaches its maximum when both populations are perfectly balanced, that is, when $p_{\mathrm{min}}=p_{\mathrm{max}}=1/2$, and vanishes when either of them is zero.

\begin{equation}
\mathcal{B}=4\:p_{\mathrm{min}}p_{\mathrm{max}}\in[0,1].
\label{eq:B}
\end{equation}

Although $\mathcal{B}$ does not encode the same information as the fidelity, it is straightforward to see that the two can be related through the populations. However, this score does not contain information about the relative phase, making it important to account for both quantities simultaneously during the optimization process.

\textbf{(iv) Non-commutativity regularizer.} In general, the Hamiltonian governing the dynamics does not commute with itself at different times. This is the principal reason why the time-ordering operator appears in the time evolution, and this strong dependence on the ordering leads to harder dynamics and more difficulties in both controlling and approximating them numerically. Accordingly, we incorporate an additional term into the objective function, namely a regularizer that penalizes the non-commutativity of the total Hamiltonian at consecutive time instants, as given in~\eqref{eq:non_comm}. This term enforces smoother temporal variations, thereby reducing the dependence of the learning process on time ordering.

\begin{equation}
\mathcal{R}(t)=\left[\pmb{\mathcal{H}}_{\mathrm{tot}}(t+\Delta t),\pmb{\mathcal{H}}_{\mathrm{tot}}(t)\right].
\label{eq:non_comm}
\end{equation}

Since Magnus propagation remains an approximation, the neural model may learn to circumvent the underlying physics and yield relatively good QFI-maximization results at the cost of driving the physical state $\ket{\Psi(t)}$ along a temporal trajectory that is not physically admissible due to the degree of freedom coming from not using the exact time-step evolution---as discussed in Section~\ref{subsec:Magnus_expansion}, this can be monitored during evaluation. To prevent the optimization from converging to such a failure mode, it is often helpful to include this regularizer with a small relative weight, thus obtaining (i) improved behavior of the Magnus propagation; (ii) more stable training, as the network is discouraged from producing rapidly oscillating or erratic controls; and (iii) greater physical interpretability, since the learned protocol more closely resembles a smooth counter-diabatic evolution rather than a highly irregular burst-like dynamics.

\textbf{Total loss function.} When defining the metric to be optimized by the model, it is important to take into account all relevant ingredients. On the one hand, for every $t\in[0,T]$, there are two principal loss terms: first, the minimization of the action through the enforcement of the Euler-Lagrange equation via~\eqref{eq:Euler-Lagrange_loss_coeffs}; and second, the non-commutativity regularizer that we just introduced. The other three terminal metrics, measured at $t=T$, are driven toward their optimal value, i.e., $\eta_{g},\:\mathrm{cos}(\Delta\varphi),\:\mathcal{B}\to 1$. For a given discrete time instant, $t_{n}$, the total loss can be defined as

\begin{equation}
\begin{aligned}
\mathcal{L}(t_{n})\;=\;
\lambda_{\mathrm{E-L}}\,\mathcal{L}_{\mathrm{E-L}}(t_n)+\lambda_{R}\,\left|\mathcal{R}(t_{n})\right|^{2}\\[4pt]
&\hspace{-4.25cm}+\delta_{n,N_t-1}\Big(\lambda_{\eta}(1-\eta_{g})^{\,2}+\lambda_{\Delta\varphi}\big(1-\cos(\Delta\varphi)\big)^{2}+\lambda_{\mathcal{B}}(1-\mathcal{B})^{2}\Big),
\end{aligned}
\label{eq:loss_at_t}
\end{equation}

\noindent where $\delta_{n,N_{t}-1}$ has been introduced to emphasize that the additional loss terms are taken into account only at the end of the evolution. The factors multiplying each loss are the weights that determine how the different terms are combined, where, in general, $\lambda_{\mathrm{E-L}}>\lambda_{\eta_{g}}=\lambda_{\mathcal{B}}>\lambda_{\Delta\varphi}>\lambda_{\mathcal{R}}$, unless stated otherwise.

The next step---although optional---follows a strategy that has been increasingly adopted in the recent PINN literature and is related to \emph{enforcing temporal causality}. According to~\cite{Wang_2024}, for losses that depend on the temporal coordinate, it is often beneficial to introduce a time-dependent weight, here denoted by $\omega(t_{n})\in[0,1]$, which modulates the loss at a given time point based on the performance of the model at all preceding time steps. In this way, and intuitively, if the network has not yet learned the earlier steps with sufficient accuracy, the loss contribution for the immediately subsequent steps remains effectively suppressed, i.e., $\omega\sim 0$. This mechanism encourages the model to respect temporal causality and to learn the underlying physics sequentially, thereby mitigating an early-training bias towards the final-time dynamics. Here, the parameter $\varepsilon_{t}$ controls the strength with which this mechanism is applied, typically taking values from $10^{-1}$ to $10^{3}$ on a logarithmic scale of base-10, depending on the problem.

\begin{equation}
\omega(t_{n})=\mathrm{exp}\left(-\varepsilon_{t}\sum_{m=0}^{n-1}\mathcal{L}(t_{m})\right).
\label{eq:causality_weight}
\end{equation}

\begin{equation}
\mathcal{L}=\frac{1}{N_{t}}\sum_{n=0}^{N_{t}-1}\omega(t_{n})\:\mathcal{L}(t_{n}).
\label{eq:total_loss}
\end{equation}

The loss defined in~\eqref{eq:total_loss} is the real-valued scalar used for the optimization of the neural model during training. Restricting the terminal metrics to $t=T$ is conceptually consistent with the enforcement of causality. However, it may become problematic if not handled carefully; until training has  sufficiently progressed, that endpoint remains effectively invisible to the network. This translated into having a nearly-zero gradient propagated through $t=T$ until the preceding time steps have been adequately learned. To this end, the value of $\varepsilon_{t}$ must be such that training is stable and should remain the same irrespective of system size to ensure meaningful comparability.

\section{Results}
\label{sec:Results}
\counterwithin*{equation}{section}
\renewcommand{\theequation}{3.\arabic{equation}}

In this section, we consider several scenarios, each defined by a Hamiltonian operator, for systems ranging, in general terms, from 2 to 6 qubits. To this end, we employ the PINN methodology introduced in Section~\ref{sec:Methodology} and illustrated in figure~\ref{fig:PINN_diagram}, whose specific features are detailed below.  In summary, this section is organized as follows: We begin by introducing the physical problems under consideration, together with the metrics used to evaluate the results. We then present the details of the experimental setup, including the general configuration adopted as well as the choice of parameters and training procedures. Next, we report the different final metrics obtained in the experiments, their learning curves, and other aspects of particular interest within our methodology. Finally, we discuss the robustness of the model with respect to network size and the issue of scalability as the number of qubits increases.

\subsection{Proposed Hamiltonian operators}
\label{subsec:proposed_H}

Regarding the experiments considered, the control parameter $g$ is interpreted as the angular frequency $\omega$ of the explicit time-dependent driving, indicating how rapidly the Hamiltonian oscillates in time and quantifying the rate of temporal modulation of the driving field or control term. In our experiments, $t\in[0,1]$ ($T=1$), and $\lambda(t)\in[0,1]$ is the interpolating scheduling function. We use $\lambda_{\mathcal{NN}}$ to denote the schedule learned by the neural network through~\eqref{eq:scheduling_function}, and $\lambda_{\mathrm{ref}}$ when referring to the non-trainable reference schedule defined in~\eqref{eq:fixed_scheduling}. The proposed control Hamiltonian is therefore defined as an interpolation between an initial static operator, $\pmb{\mathcal{H}}_{\omega}^{\mathrm{initial}}$, and a final time-dependent operator, $\pmb{\mathcal{H}}_{\omega}^{\mathrm{final}}(t)$, given in~\eqref{eq:Homega_initial} and~\eqref{eq:Homega_final}, respectively.

\begin{equation}
\pmb{\mathcal{H}}_{\omega}^{\mathrm{initial}}=\sum_{i=1}^{q}h\:\sigma_{\mathrm{X}}^{(i)},
\label{eq:Homega_initial}
\end{equation}
\begin{equation}
\pmb{\mathcal{H}}_{\omega}^{\mathrm{final}}(t)=-\mathrm{sin}(\omega t)\sum_{i\neq j}\frac{1}{|i-j|^{\alpha}}\sigma_{\mathrm{X}}^{(i)}\sigma_{\mathrm{Y}}^{(j)}+\mathrm{cos}(\omega t)\sum_{i=1}^{q}\sigma_{\mathrm{Z}}^{(i)}.
\label{eq:Homega_final}
\end{equation}

In our setup, $h$ can be understood as the strength of the local single-qubit transverse field, that is, the transverse-field amplitude, while $\alpha$ is the power-law exponent that governs how the interaction strength decays with the distance between qubits. For all the cases, $h=1$ is kept fixed, and the frequency is likewise set to $\omega=1$, with the aim that these numerical results may serve as a benchmark. As for $\alpha$, it is used to distinguish among three different operators: (i) $\alpha\to\infty$, corresponding to the nearest-neighbor Hamiltonian, in which interactions are restricted to immediately adjacent qubits; (ii) $\alpha=3$, which is used for dipolar interactions of the form $1/r^{3}$; and (iii) $\alpha=6$, which is used for van der Waals interactions in Rydberg atoms. In the special case $\alpha\to\infty$, the factor $1/|i-j|^{\alpha}$ reduces to the nearest-neighbor interaction strength $J$, which is likewise fixed to $J=1$ in all scenarios.

\begin{equation}
\pmb{\mathcal{H}}_{\omega}(t)=(1-\lambda(t))\:\pmb{\mathcal{H}}_{\omega}^{\mathrm{initial}}+\lambda(t)\:\pmb{\mathcal{H}}_{\omega}^{\mathrm{final}}(t).
\label{eq:Hamiltonian_proposed}
\end{equation}

\begin{equation}
\pmb{\mathcal{H}}_{\mathrm{tot}}(t)=\pmb{\mathcal{H}}_{\omega}(t)+\frac{\mathrm{d}\lambda}{\mathrm{d}t}\pmb{\mathcal{A}}_{\lambda}(t).
\label{eq:final_Hamiltonian}
\end{equation}

In this way, the set $\{\lambda,\pmb{\mathcal{A}}_{\lambda}\}$ constitutes the trainable degrees of freedom that the model can learn and modulate to maximize the QFI while satisfying the underlying physics. Finally, to compute the QFI during training, we use central finite differences to evaluate $\ket{\partial_{\omega}\Psi(T)}$, with $\delta\omega=10^{-10}\omega$. Moreover, since this calculation is extremely sensitive, double-precision floating-point arithmetic is employed.\footnote{In practice, \emph{float128} precision is used, since we generally work with complex numbers: 64 bits for the real part and a further 64 bits for the imaginary part.}

\subsection{Evaluation metrics}
\label{subsec:eval_metrics}

In our experiments, no ground-truth or reference solution is available against which the model prediction can be compared in order to directly assess the accuracy of the numerical approximations. Consequently, we rely on a set of physically motivated evaluation metrics to verify that the obtained results correspond to a valid physical solution. In addition to the metrics already introduced---namely, the normalized QFI, the fidelity,  and the state extremal-balance score---, the quality of the learned dynamics is further assessed through two complementary quantities. First, we evaluate the normalized Schrödinger residual $\varepsilon_{\mathrm{Schr}}$~\eqref{eq:sch_metric}, which measures how accurately the predicted evolution satisfies the time-dependent Schrödinger equation under the total Hamiltonian $\pmb{\mathcal{H}}_{\mathrm{tot}}$ throughout the protocol. More precisely, this quantity corresponds to a global $L^{2}$-in-time residual normalized by the associated Hamiltonian action, so that small values indicate that the propagated state remains close to the physical dynamics enforced during training. Second, we evaluate the unitary error $\varepsilon_{\mathrm{uni}}$ of the block propagators used in the Magnus evolution~\eqref{eq:uni_error}. This metric quantifies the root-mean-square deviation of the window propagators from the exact unitary condition $\pmb{\mathcal{U}}^{\dag}\pmb{\mathcal{U}}=\pmb{I}$, measured by the Frobenius norm and normalized by the Hilbert-space dimension through the factor $2^{q}$. Since exact quantum evolution must preserve unitarity, values of $\varepsilon_{\mathrm{uni}}$ close to zero indicate that the numerical time propagation remains stable and physically consistent.

\begin{equation}
\varepsilon_{\mathrm{Schr}}\equiv\left(\frac{\int_{0}^{T}||i\partial_{t}\ket{\Psi(t)}-\pmb{\mathcal{H}}_{\mathrm{tot}}\ket{\Psi(t)}||^{2}_{2}\:dt}{\int_{0}^{T}||\pmb{\mathcal{H}}_{\mathrm{tot}}\ket{\Psi(t)}||^{2}_{2}}\right)^{1/2}.
\label{eq:sch_metric}
\end{equation}

\begin{equation}
\varepsilon_{\mathrm{uni}}\equiv\left(\frac{1}{n_{w}2^{q}}\sum_{w=1}^{n_{w}}\left|\left|\pmb{\mathcal{U}}^{(w)}\pmb{\mathcal{U}}^{(w)}-\pmb{I}\right|\right|^{2}_{F}\right)^{1/2}.
\label{eq:uni_error}
\end{equation}

Altogether, these metrics provide a direct validation that the PINN outputs not only optimize the target objective, but also generate trajectories compatible with the fundamental dynamical constraints of the problem.

\subsection{Standard configuration}
\label{subsec:standard_config}

The methodology we propose involves several hyperparameters that can be adjusted and fixed even before training the neural model. It is therefore essential to establish a ``standard configuration'' that will serve as the baseline throughout all numerical analyses, unless otherwise specified. The parameters described below are the outcome of an extensive comparative numerical performance study based on grid search, with Optuna~\cite{optuna_2019} also employed as a complementary methodology.\\

\begin{table*}[!b]
\centering
\normalsize
\setlength{\tabcolsep}{4pt}
\renewcommand{\arraystretch}{1.12}

\newcommand{\ann}[1]{\text{\scriptsize #1}}

\caption{\textbf{Summary of PINN performance across system size.} For the different Hamiltonians considered, and for each number of qubits $q$, we report the achieved QFI efficiency $\eta_{\omega}$, final-state fidelity $\mathcal{F}$, and final-state extremal-balance score $\mathcal{B}$ (mean $\pm$ std), together with the final Euler-Lagrange residual $\mathcal{L}_{\mathrm{E-L}}$ and the uncertainty made by the Magnus expansion on the QFI, $\varepsilon\left(\eta_{\omega}\right)$. The values shown in parenthesis on the right of each reported metric indicate the relative improvement of the corresponding metric with respect to the reference solution.}
\label{tab:results_summary}

\begin{tabular*}{\textwidth}{@{\extracolsep{\fill}}@{} c |
S[table-format=1.3(3),table-align-uncertainty] @{\;} l
S[table-format=1.3(3),table-align-uncertainty] @{\;} l
S[table-format=1.3(3),table-align-uncertainty] @{\;} l
| S[parse-numbers=false] S[parse-numbers=false] @{}}
\toprule
\(q\) &
\multicolumn{2}{c}{\(\eta_{\omega}\)} &
\multicolumn{2}{c}{\(\mathcal{F}\)} &
\multicolumn{2}{c}{\(\mathcal{B}\)} &
\multicolumn{1}{c}{\(\mathcal{L}_{\mathrm{E-L}}\)} &
\multicolumn{1}{c}{\(\varepsilon\!\left(\eta_{\omega}\right)\)} \\
\midrule

\multicolumn{9}{@{}l}{\textbf{Hamiltonian: Nearest neighbors}}\\
\specialrule{0.05pt}{0pt}{4pt}
2 & 0.940(10) & \ann{(+20\%)}   & 0.706(8)  & \ann{(\(\times 2.5\))}   & 0.910(3) & \ann{(+24\%)}    & {$2\times 10^{-4}$} & {$9\times 10^{-6}$} \\
3 & 0.521(4)  & \ann{(\(\times 3\))} & 0.827(3)  & \ann{(\(\times 330\))}  & 0.725(4) & \ann{(\(\times 9\))}  & {$1\times 10^{-3}$} & {$4\times 10^{-4}$} \\
4 & 0.918(3)  & \ann{(\(\times 3.5\))} & 0.898(5)  & \ann{(\(\times 23\))}   & 0.911(1) & \ann{(\(\times 10\))} & {$8\times 10^{-5}$} & {$1\times 10^{-5}$} \\
5 & 0.932(7)  & \ann{(\(\times 7\))} & 0.957(8)  & \ann{(\(\times 24\))}   & 0.954(6) & \ann{(\(\times 490\))} & {$1\times 10^{-4}$} & {$4\times 10^{-4}$} \\
6 & 0.984(6)  & \ann{(\(\times 9\))} & 0.990(7)  & \ann{(\(\times 32\))}   & 0.990(6) & \ann{(\(\times 900\))} & {$1\times 10^{-5}$} & {$2\times 10^{-5}$} \\
\midrule

\multicolumn{9}{@{}l}{\textbf{Hamiltonian: Dipolar}}\\
\specialrule{0.05pt}{0pt}{4pt}
2 & 0.941(0.009)  & \ann{(+36\%)}   & 0.706(0.007)  & \ann{(\(\times 3\))}  & 0.910(0.007) & \ann{(+43\%)}    & {$2\times 10^{-4
}$} & {$3\times 10^{-5
}$} \\
3 & 0.520(0.003)  & \ann{(\(\times 3\))} & 0.773(0.004)  & \ann{(\(\times 3\))} & 0.745(0.002) & \ann{(\(\times 11\))} & {$1\times 10^{-3}$} & {$3\times 10^{-4}$} \\
4 & 0.776(0.050) & \ann{(\(\times 3\))} & 0.572(0.003)  & \ann{(\(\times 2\))} & 0.851(0.003) & \ann{(\(\times 9\))} & {$6\times 10^{-4}$} & {$8\times 10^{-5}$} \\
5 & 0.868(0.060) & \ann{(\(\times 8\))} & 0.939(0.007) & \ann{(\(\times 18\))}   & 0.948(0.003) & \ann{(\(\times 700\))} & {$2\times 10^{-4}$} & {$9\times 10^{-5}$} \\
6 & 0.983(0.001) & \ann{(\(\times 9\))} & 0.989(0.001)  & \ann{(\(\times 25\))}  & 0.989(0.001) & \ann{(\(\times 900\))} & {$1\times 10^{-5}$} & {$4\times 10^{-5}$} \\
\midrule

\multicolumn{9}{@{}l}{\textbf{Hamiltonian: Trapped ions}}\\
\specialrule{0.05pt}{0pt}{4pt}
2 & 0.941(9)  & \ann{(+21\%)}   & 0.707(8)  & \ann{(\(\times 3.4\))}  & 0.910(1) & \ann{(+40\%)}    & {$2\times 10^{-4}$} & {$1\times 10^{-5}$} \\
3 & 0.559(2)  & \ann{(\(\times 2.5\))} & 0.859(4)  & \ann{(\(\times 250\))} & 0.824(2) & \ann{(\(\times 44\))} & {$1\times 10^{-3}$} & {$2\times 10^{-4}$} \\
4 & 0.810(50) & \ann{(\(\times 2.5\))} & 0.574(4)  & \ann{(\(\times 2.8\))} & 0.881(1) & \ann{(\(\times 24\))} & {$6\times 10^{-4}$} & {$4\times 10^{-5}$} \\
5 & 0.921(50) & \ann{(\(\times 6\))} & 0.937(10) & \ann{(\(\times 7\))}   & 0.942(5) & \ann{(\(\times 56\))} & {$1\times 10^{-4}$} & {$3\times 10^{-4}$} \\
6 & 0.960(50) & \ann{(\(\times 6\))} & 0.989(8)  & \ann{(\(\times 96\))}  & 0.987(7) & \ann{(\(\times 430\))} & {$3\times 10^{-5}$} & {$6\times 10^{-5}$} \\
\bottomrule
\end{tabular*}

\vspace{4pt}
\end{table*}

\textbf{Loss function.} As regards the parameters of the loss function, and in particular the weights assigned to each constituent term, we impose the hyerarchy $\lambda_{\mathrm{E-L}}>\lambda_{\eta_{g}}=\lambda_{\mathcal{B}}>\lambda_{\Delta\varphi}>\lambda_{\mathcal{R}}$, reflecting the physical bias associated with the relative importance of the physics enforced by each contribution. In practice, this set will be fixed to $\{\lambda_{\mathrm{E-L}},\lambda_{\eta_{g}},\lambda_{\mathcal{B}},\lambda_{\Delta\varphi},\lambda_{\mathcal{R}}\}=\{10^{3},1,1,10^{-1},10^{-2}\}$. On the other hand, as regards causality enforcement, we set $\varepsilon_{t}=1$, which does not constitute an excessive modification of the vanilla methodology, while still being a non-trivial alteration of the function.

\textbf{Time sampling and Magnus propagation.} Although this has already been specified before, it is worth stating explicitly that, for the temporal discretization, we use $N_{t}=2^{8}$ equally spaced time steps over the interval $[0,T]$ with $T=1$, which are kept fixed throughout the entire training process.\footnote{Even though recent PINN literature has proposed techniques that relocate training points according to certain criteria, this is not the approach adopted here. Our aim is to assess the intrinsic approximation capability of the PINN primarily through a well-defined loss function.} For the Magnus expansion, we consider terms up to order $p=3$ and employ $n_{w}=2^{4}$ windows, with, therefore, $m=2^{4}$ temporal points in each of them. This yields, on average, an error of approximately $\sim 10^{-4}$ in the time evolution.

\textbf{Neural architecture.} Regarding the network parameters, we have, in broad terms, employed a fairly basic architecture consisting of two independent multilayer perceptrons (MLPs) (see figure~\ref{fig:PINN_diagram}). The MLP that predicts the function $\lambda(t)$ has a fixed architecture of 3 layers with 50 neurons each, whereas the MLP that gives the AGP coefficients, $\pmb{\mathcal{A}}_{\lambda}(t)$, consists of 6 layers, likewise with 50 neurons per layer. The activation function used in all hidden layers is SiLU, while the outputs are left without any activation, so that they can vary over the physically required range. Finally, the trainable layer parameters (weights and biases) are initialized according to the Xavier uniform criterion with gain 1.

\textbf{Training details.} The optimizer used in all experiments is Shampoo with Adam in the Preconditioner's eigenbases (SOAP)~\cite{SOAP_2025}, which has only recently been introduced in the literature and has proven to be substantially competitive in the context of PINNs~\cite{SOAP_PINNs}. It is worth noting that, although SOAP is somewhat slower than other optimizers such as Adam, this is compensated by the generally smaller number of iterations required to converge to the true physical solution. Accordingly, all experiments are run for a total of 25,000 epochs, with a fixed learning rate of $10^{-4}$, without the need to implement any additional scheduler. All training was performed in PyTorch on an NVIDIA A100 (40GB) GPU.

\subsection{Numerical simulations}

In this section, we present the main results obtained, together with the conclusions drawn from the full set of analyses performed. To this end, all experiments make use of the same neural architecture, whose specific features, along with those of the loss function, have been presented in the preceding sections, while the Hamiltonian dynamics considered corresponds to the three cases introduced in Section~\ref{subsec:proposed_H}. In addition, for purely comparative purposes, we define the corresponding ``reference'' prediction for each particular case study as the outcome of training the same network under the same training settings, but with weights $\lambda_{\mathrm{E-L}}=10^{3}$ and $\lambda_{\eta_{\omega}}=\lambda_{\mathcal{B}}=\lambda_{\Delta\varphi}=\lambda_{\mathcal{R}}=0$; that is, only the Euler-Lagrange loss remains active, while no Fisher-information maximization is enforced. In this way, these results provide a reliable baseline against which any improvement in information performance can be compared.

\textbf{Model performance across system size.} Table~\ref{tab:results_summary} reports the different physical metrics of interest for the various dynamics and size of the system, together with the physical loss $\mathcal{L}_{\mathrm{E-L}}$ obtained at the end of training and the error $\varepsilon(\eta_{\omega})$ incurred when comparing the final QFI calculated by Magnus propagation with that obtained with an exact sequential time propagation. In general terms, we see a consistent and physically coherent picture of the PINN performance across the three Hamiltonian dynamics. For most system sizes, the method achieves high values of QFI efficiency, fideity and extremal-balance score, while simultaneously keeping the final Euler-Lagrange residual at low values and the Magnus-based uncertainty on the QFI under control. Altogether, this indicates that the network is not only increasing the metrological figure of merit, but is doing so while preserving the underlying dynamical restrictions imposed by the variational formulation. A second feature is the clear improvement with respect to the corresponding reference solutions, which confirms that the enforcement of the additional metrological objectives produces a substantial gain over a purely Euler-Lagrange-driven baseline. The table also reveals that this behavior is not strictly monotonic with system size and that certain small-size configurations deviate from that favorable trend.

The most evident anomalous case is $q=3$, which is systematically associated with poorer values of $\eta_{\omega}$, $\mathcal{F}$, and $\mathcal{B}$, together with larger residual losses, and a similar, though milder, effect is also seen for $q=4$ for the dipolar and trapped-ion models. A detailed analysis of this anomalous regime $q=3$, including the role of extremal-subspace tracking and symmetry mismatch, is provided in~\ref{sec:appendix3}. In brief, these cases can be interpreted as particularly unfavorable finite-size regimes from the point of view of symmetry and dynamical accessibility of the optimal metrological subspace.

\begin{figure}[!t]
    \centering
    \includegraphics[width=1.0\columnwidth]{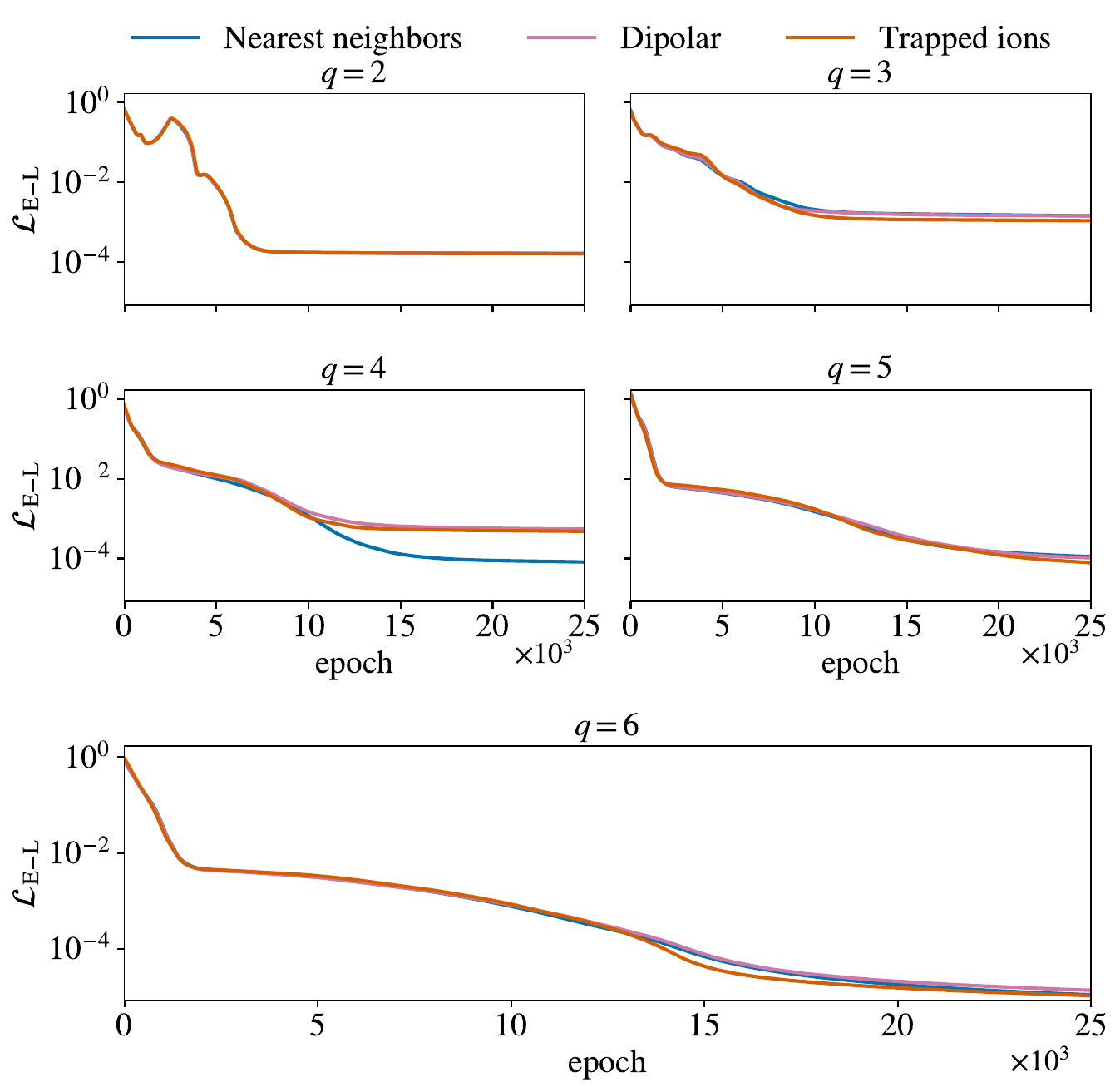}
    \caption{\justifying Euler-Lagrange loss $\mathcal{L}_{\mathrm{E-L}}$ versus training epoch for $q=2-6$ qubits, comparing Hamiltonians (one panel per system size).}
    \label{fig:losses}
\end{figure}

\textbf{Learning curves.} Figure~\ref{fig:losses} shows that the Euler-Lagrange loss decreases consistently during training for all dynamics and for all system sizes, which indicates that the PINN is able to learn physically admissible gauge-potential dynamics in a stable way. In most cases, there is a sharp initial drop, followed by a slower decay toward a plateau at smaller residual values, suggesting that the model first captures the dominant structure of the solution and then refines it more gradually. In general, the curves for $q=2,5,6$ are comparatively lower and end with smaller losses, so these scenarios appear to be easier to satisfy within the same training time. In particular, for $q=6$, and to a lesser extent for $q=5$, the loss still appears to decrease by the time training reaches 25,000 epochs. For consistency with the other experiments, however, training is assumed to stop at that point, since the observed improvement is only marginal.

By contrast, the cases $q=3$, and to a lesser extent $q=4$ for the dipolar and trapped-ion models, remain systematically at higher residual levels throughout the training, in agreement with the behavior already discussed in the previous section. As discussed in the previous section and analyzed in more detail in~\ref{sec:appendix3}, these less favorable cases are better interpreted as finite-size regimes with a more restrictive dynamical structure, where the evolution has greater difficulty remaining within the metrologically optimal sectors. This makes the corresponding learning problem more demanding and leads to larger final $\mathcal{L}_{\mathrm{E-L}}$ residuals.

\begin{figure}[!t]
    \centering
    \includegraphics[width=1.0\columnwidth]{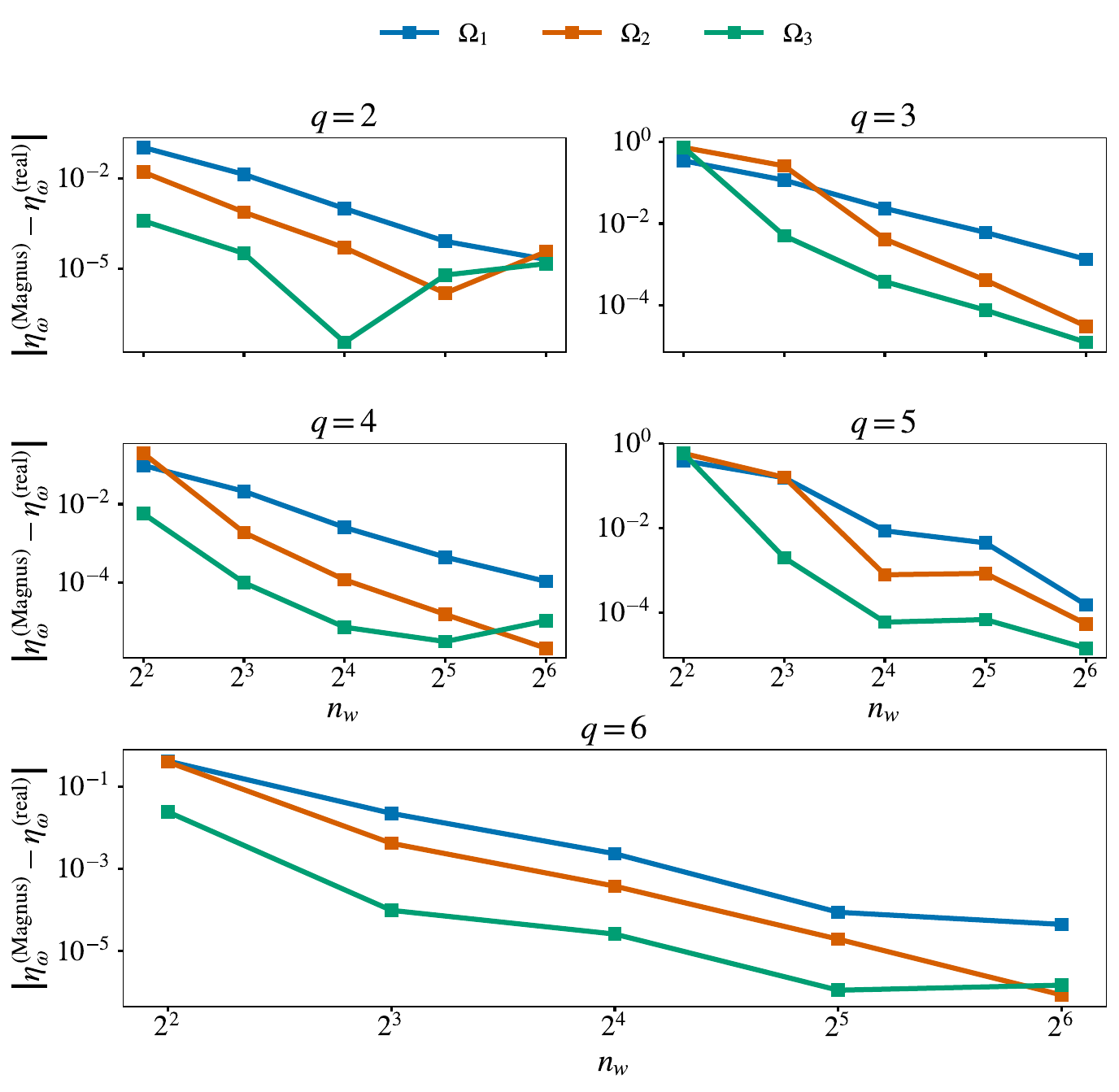}
    \caption{\justifying Log-scale error in the normalized QFI between driving the entire time evolution in a sequential way with respect to using the Magnus expansion approximation---up to third order---, versus several number of windows $n_{w}$ for $q\in\{2,3,4,5,6\}$ qubits. The Hamiltonian of nearest-neighbors has been considered.}
    \label{fig:QFI_error}
\end{figure}

\textbf{Error of the time evolution.} Figure~\ref{fig:QFI_error} shows the variation of the experimental error, measured as the absolute difference, induced by the measurement of the normalized QFI when Magnus propagation is used, relative to the sequential time evolution of the state of the system $\ket{\Psi(t)}$. For this numerical analysis, we consider system sizes ranging from $q=2$ to $q=6$ qubits, while varying the number of windows $n_{w}$ from $2^{2}$ to $2^{6}$, with the total number of temporal instants kept fixed at $N_{t}=2^{8}$. For this study, we consider up to three orders ($p=3$) in the Magnus approximation, following the terms described in~\ref{sec:appendix1}.

As can be seen, the overall behavior is consistent with the theoretical trend presented in Section~\ref{subsec:Magnus_expansion}: Using $2^{4}$ windows and $p=3$, the resulting error remains on the order of $10^{-4}$ across all scenarios. Although this level of error can be sufficient for our experiments, it is nevertheless of interest to verify this trend empirically in cases where higher accuracy may be required. It should also be emphasized that the dynamics considered here correspond only to the nearest-neighbor case, so certain mildly anomalous behaviors that deviate from the general trend---such as the $p=3$ case for $q=2$, where the trend seems to break down for $n_{w}>2^{4}$---may be attributable to the specific set of learned interactions between qubits. Despite this, a similar behavior should be expected in the remaining dynamics, since the theoretical expression for this error is independent of the particular dynamics considered (equation~\eqref{eq:theo_error_Magnus_expansion}.

\begin{figure}[!t]
    \centering
    \includegraphics[width=1.0\columnwidth]{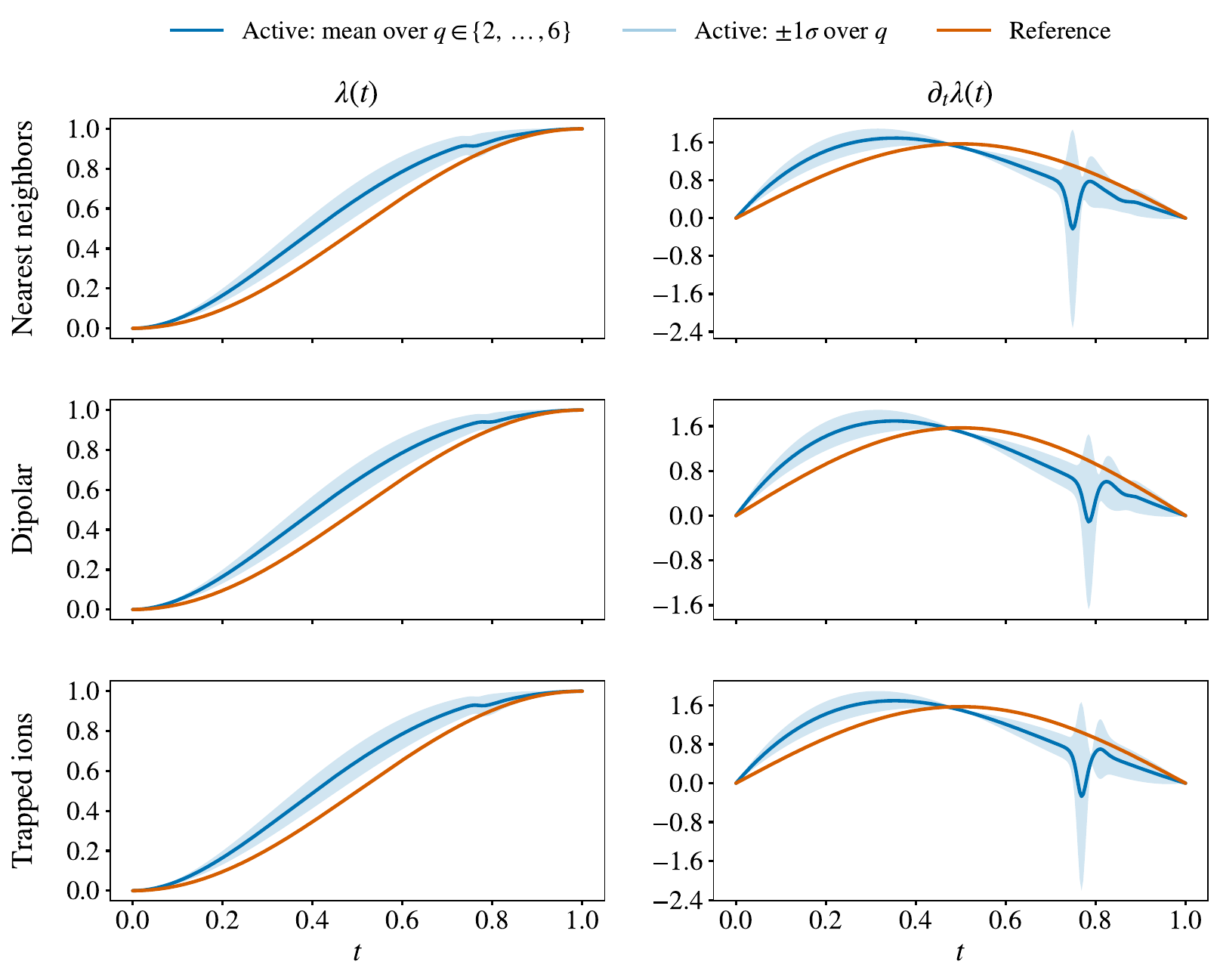}
    \caption{\justifying Mean $\lambda(t)$ and $\partial_{t}\lambda(t)$ trajectories for different Hamiltonians (rows), averaged over $q=2-6$ qubits (solid). Shaded bands indicate the standard deviation across qubit counts, and the non-trainable scheduling function used as reference is shown for comparison (orange).}
    \label{fig:lambda}
\end{figure}

\begin{figure*}[!b]
    \centering
    \includegraphics[width=2.0\columnwidth]{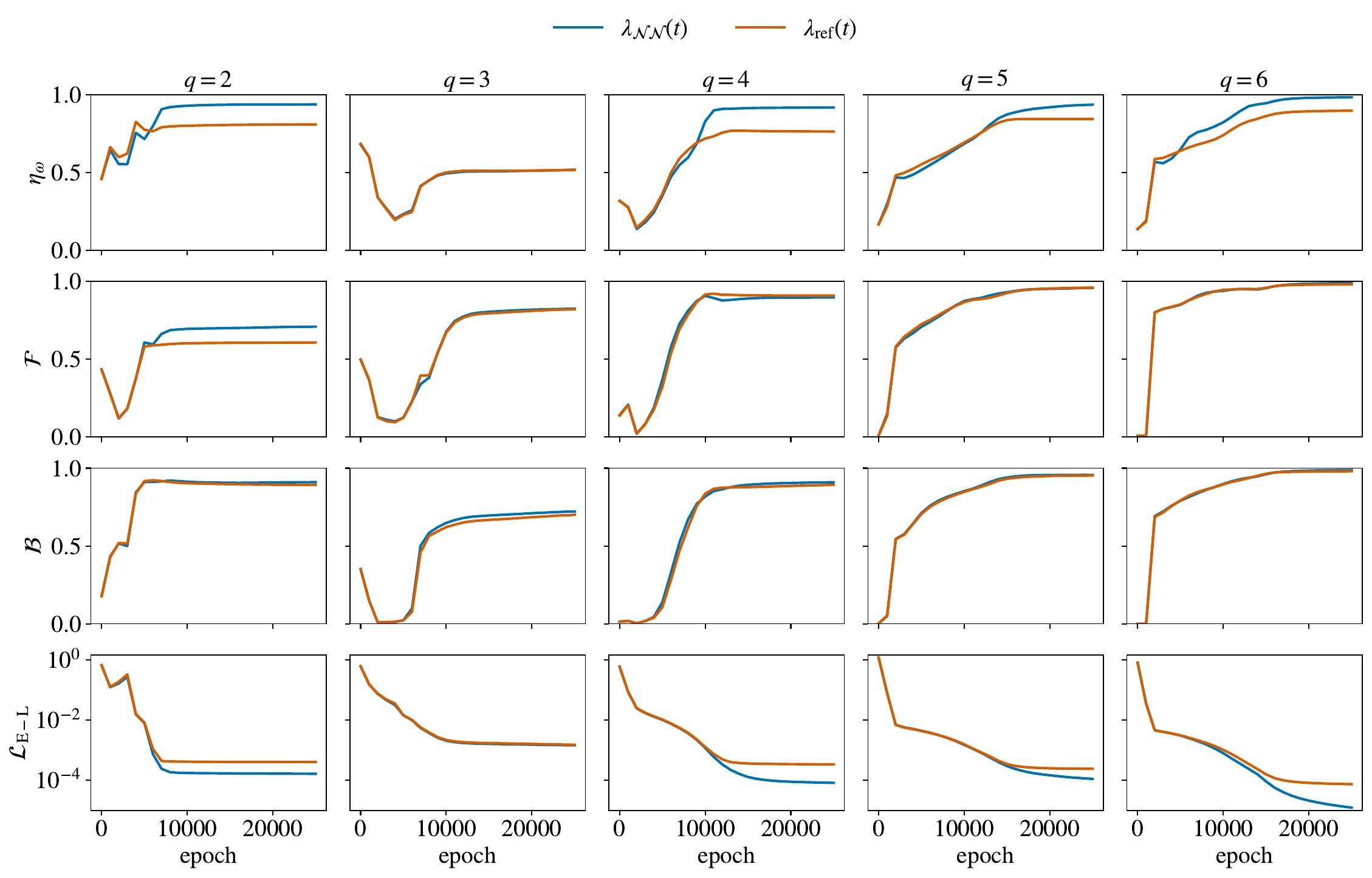}
    \caption{\justifying Training trajectories of the final metrics $\eta_{\omega}$, $\mathcal{F}$, and $\mathcal{B}$ for $q=2-6$ qubits. The Euler-Lagrange loss is also considered for comparison purposes. Curves compare a trainable time-dependent coupling $\lambda_{\mathcal{NN}}(t)$ against the fixed reference $\lambda_{\mathrm{ref}}(t)$. The Hamiltonian of nearest-neighbors has been considered.}
    \label{fig:metrics_vs_lambda}
\end{figure*}

\textbf{Impact of the learned scheduling function $\lambda(t)$.} As described in Section~\ref{subsec:Scheduling_function} of the methodology, not only can the coefficients of the basis in which the AGP can be decomposed---and, consequently, the $\pmb{\mathcal{A}}_{\lambda}(t)$ operator itself---be learned by the network, but the scheduling function $\lambda(t)$ itself may also be adjusted as an additional degree of freedom for the PINN. As recalled above, once the control Hamiltonian $\pmb{\mathcal{H}}_{\omega}(t)$ has been defined in~\eqref{eq:Hamiltonian_proposed} as a driving from an initial operator that is ``easy'' to implement experimentally to a given final operator, the function $\lambda(t)$ plays an interpolating role and must therefore remain confined to the interval $[0,1]$\footnote{It is worth noting that, mathematically, $\lambda(t)$ may exceed 1 at certain times. Physically, this does not necessarily make the protocol infeasible: in our case, $\lambda(t)>1$ would mean that the transverse field changes sign while the interaction term is further strengthened. This remains admissible as long as the resulting parameters stay within hardware limits, although it would correspond to optimizing over an extended control landscape rather than along the original fixed path.}. Figure~\ref{fig:lambda} shows the learned $\lambda(t)$ for the different Hamiltonian dynamics considered. The mean is represented by a solid line, while a shaded band indicates de  $1\sigma$ deviation across all system sizes considered, from 2 to 6 qubits. For comparison, the non-trainable ``reference'' $\lambda(t)$, given in~\eqref{eq:fixed_scheduling}, is also plotted as a solid line for comparison purposes. This function, trained explicitly according to the methodology introduced in Section~\ref{subsec:Scheduling_function}, is seen to remain broadly similar to the reference profile while showing clear deviations from it. These differences become particularly evident in the derivative, which directly multiplies the potential $\pmb{\mathcal{A}}_{\lambda}(t)$ and determines both the rate of the transition and the strength with which these terms enter the total Hamiltonian.

\begin{figure*}[!htbp]
    \centering
    \includegraphics[width=2.0\columnwidth]{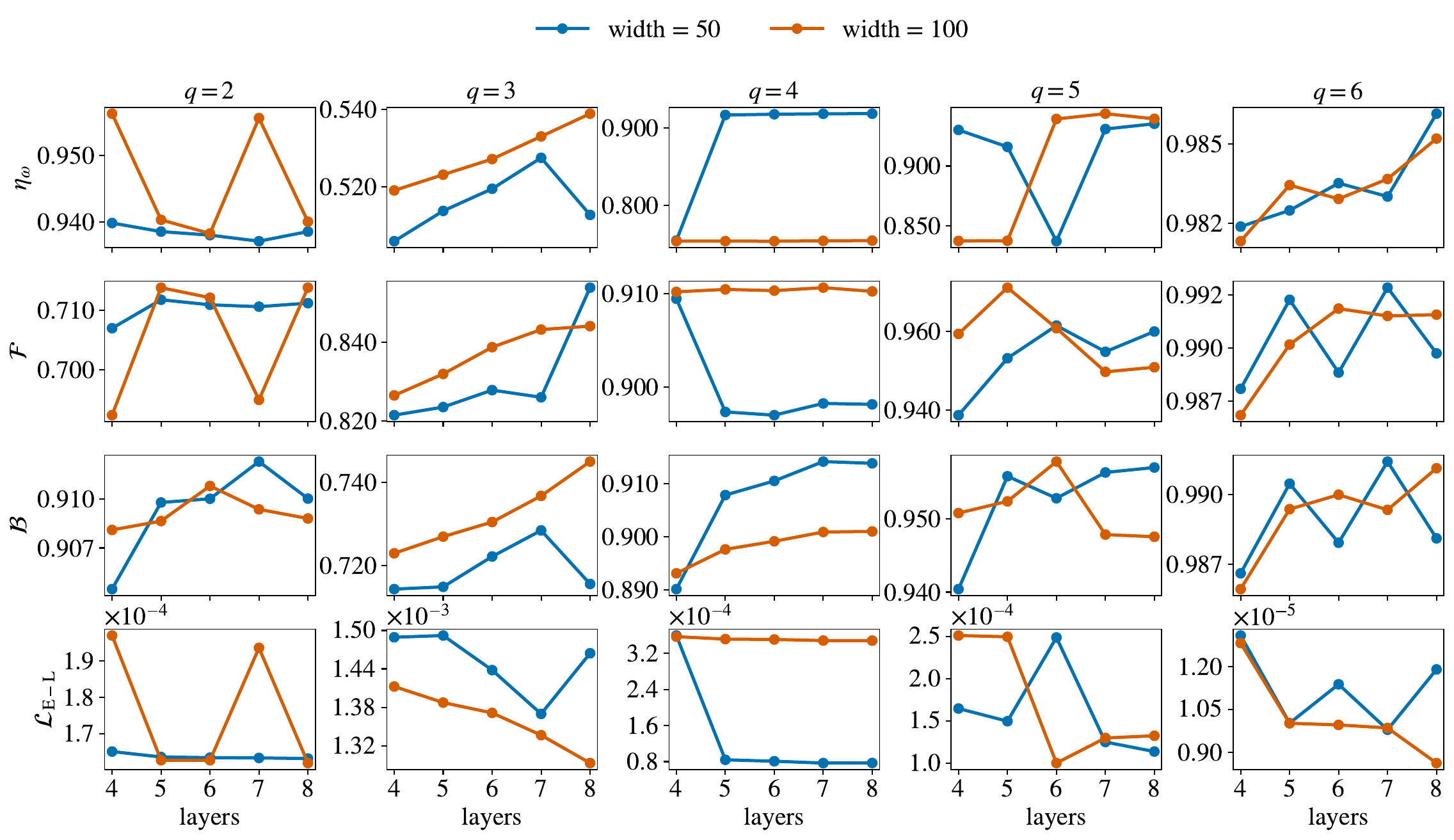}
    \caption{\justifying Performance metrics $\eta_{\omega}$, $\mathcal{F}$, and $\mathcal{B}$ as functions of the number of layers considered for the PINN for $q=2-6$ qubits. Each panel compares the two widths considered, showing how neural size can affect the final model performance across system sizes. The Hamiltonian of nearest-neighbors has been considered.}
    \label{fig:metrics_vs_size}
\end{figure*}

Whereas the reference protocol prescribes a gradually increasing function in $t\in[0,T]$, with a maximal derivative at the midpoint, the PINN learns an alternative profile that remains smooth but develops a pronounced pulse toward the end of the evolution, around $t\approx 0.8$. The emergence of this late-time pulse suggests that the PINN allocates a stronger counter-diabatic correction to the stage of the evolution most susceptible to diabatic leakage. This behavior is consistent with a reduced effective spectral gap, requiring enhanced control to suppress non-diabatic transitions and preserve adiabatic following. It is worth emphasizing that this learned $\lambda(t)$ is constrained by the methodology adopted here; different results could emerge if the proposed hard enforcement were modified or if values $\lambda>1$ were allowed. Even so, it is particularly noteworthy that the model identifies an alternative scheduling function to the one most commonly employed in the literature about CD protocols, although such a result is likely to remain strongly problem dependent.

Nevertheless, it is important not only to compare the learned scheduling function, $\lambda_{\mathcal{NN}}(t)$, with the reference profile, $\lambda_{\mathrm{ref}}(t)$, but also to examine whether this difference actually translates into improved performance of the learned dynamics. Figure~\ref{fig:metrics_vs_lambda} presents the evolution, as a function of the number of training epochs, of the final metrics $\eta_{\omega}$, $\mathcal{F}$, and $\mathcal{B}$ for different system sizes, together with the corresponding loss $\mathcal{L}_{\mathrm{E-L}}$. In each case, a distinction is made between the use of the reference schedule $\lambda_{\mathrm{ref}}(t)$ and the setting in which $\lambda(t)$ is treated as a trainable function. This analysis has been carried out specifically for the nearest-neighbor Hamiltonian, which is used here as a representative example. With the exception of the $q=3$ case, which has already been discussed, one may observe that the QFI generally increases substantially when the scheduling function is made trainable, rising, for instance, from $\eta_{\omega}\sim 0.8$ to $\eta_{\omega}\sim 1.0$ in the cases of 2 and 4 qubits. By contrast, the metrics $\mathcal{F}$ and $\mathcal{B}$ do not vary substantially between the two cases. Even so, the physical action loss is generally reduced when the scheduling function is made trainable, including for $q=3$, where the improvement, though modest, remains discernible. Overall, these results indicate that incorporating this additional degree of freedom into the PINN is highly beneficial for addressing the problem under consideration.

In terms of scalability, we have introduced a parallel branch in the model, consisting of 4 layers with 50 neurons each, in order to predict $\lambda(t)$. Nevertheless, it would be worth investigating whether a single MLP could simultaneously learn an adequate scheduling function together with the AGP. In any case, the increase in the number of parameters is not expected to account for the bulk of the computational overhead, since the principal scalability bottleneck lies in the exponential growth with $q$, rather than in these few additional layers.

\textbf{Robustness of the model versus neural size.} From a methodological standpoint, it is important to examine how the model behaves in terms of performance as a function of the network size considered, in order to determine whether a clear positive trend exists with respect to the number of parameters. To investigate this, we vary the depth and width of the MLP within the PINN that predicts the coefficients of the decomposition of $\pmb{\mathcal{A}}_{\lambda}(t)$, considering architectures ranging from 4 to 8 layers and 50 to 100 neurons per layer. In this way, we span configurations from relatively shallow and narrow networks to deep and wide ones, together with several intermediate cases. Figure~\ref{fig:metrics_vs_size} presents the different final evaluation metrics considered, together with the loss $\mathcal{L}_{\mathrm{E-L}}$, across these scenarios.

As can be seen, all variables respond favorably, in broad terms, to an increase in network width, which is to be expected given the substantial enhancement in the expressive capacity of the model. However, the differences are not particularly pronounced, as the results already lie in a generally strong regime; indeed, the largest variations are of the order of $\sim 10\%$, as in the $q=4$ case, which is perhaps the most noticeable. In terms of the depth of the network, measured by the number of layers, the model generally benefits from increasing it, as reflected in a positive trend in the evaluation metrics and a negative trend in the loss. This is again consistent with the improved predictive capacity associated with deeper architectures. Nevertheless, it should be emphasized that these differences remain relatively small. Far from being a drawback, this is in fact a positive result, since it supports the conclusion that our PINN is robust with respect to architectural variations and that any improvements, when present, are not substantial. It is important to note that the example Hamiltonian considered here is the nearest-neighbor one, although similar behavior may be expected for any other dynamics without compromising the generality of our conclusions. Indeed, while the absolute predictive performance of the model may depend on the specific dynamics under consideration---since some cases may be intrinsically more difficult to solve than others---in relative terms, when different configurations are compared within the same dynamics, the performance is expected to improve or deteriorate according to the number of model parameters and, therefore, its predictive capacity.

\begin{figure*}[!t]
    \centering
    \includegraphics[width=2.0\columnwidth]{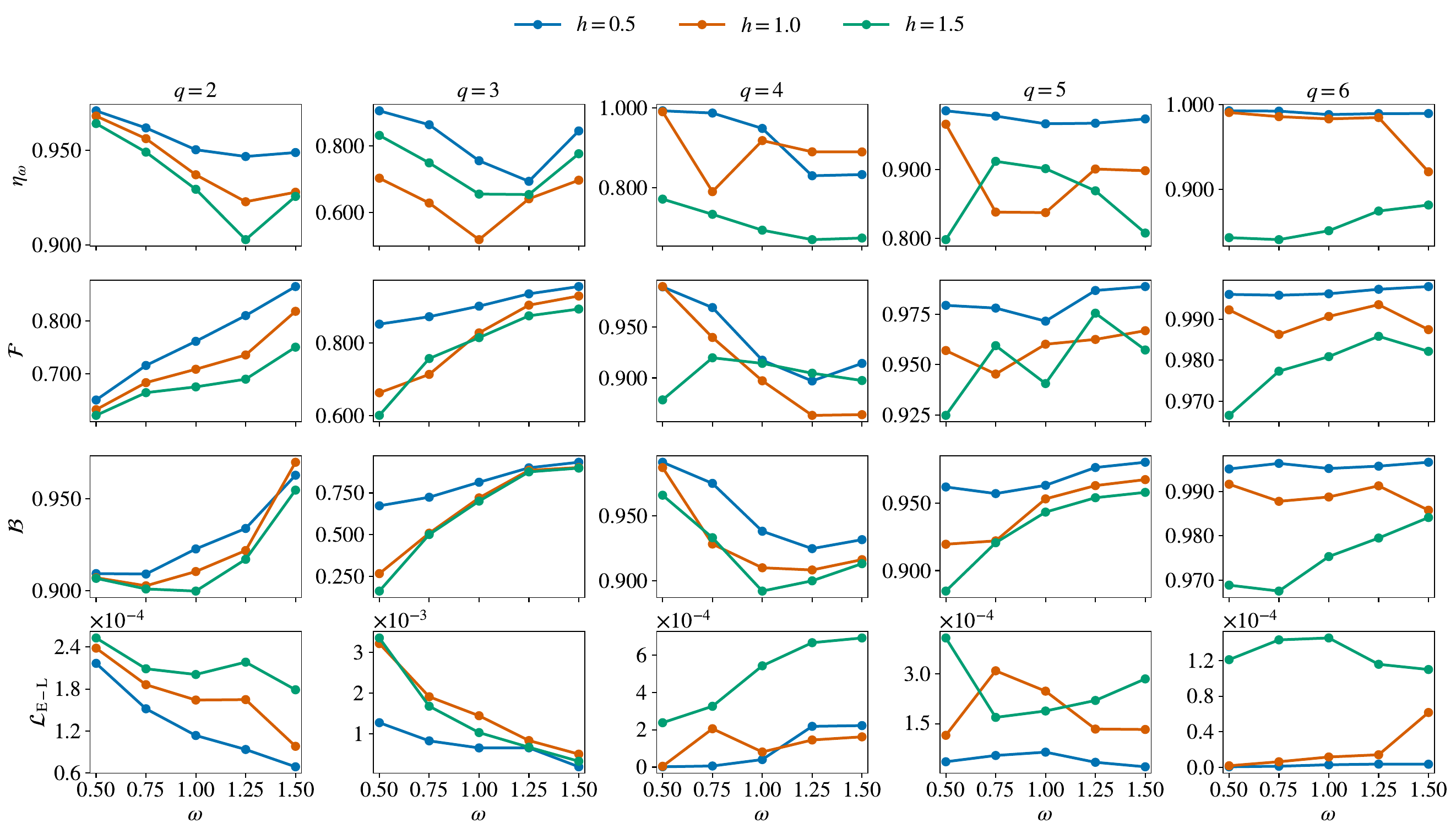}
    \caption{\justifying Variation of the final metrics $\eta_{\omega}$, $\mathcal{F}$, and $\mathcal{B}$ as a function of the driving frequency $\omega$, for different system sizes $q$. Each panel shows one value of $q$, while different curves correspond to different values of the field strength $h$. The Hamiltonian of nearest-neighbors has been considered.}
    \label{fig:figure__metrics_vs_omega_and_h}
\end{figure*}

\textbf{Robustness of the model versus physical parameters.} Up to this point, all analyses have been carried out for the Hamiltonian operators described in Section~\ref{subsec:proposed_H}, with both the frequency $\omega=1$ and the field-strength constant $h=1$ kept fixed. The values obtained for the different metrics under this configuration are those reported in the corresponding table~\ref{tab:results_summary}. However, a natural question is how sensitive the model performance may be to these parameters and whether the proposed PINN can therefore be regarded as robust with respect to the underlying physics of the problem. Figure~\ref{fig:figure__metrics_vs_omega_and_h} presents the metrics of interest, but now the frequency is varied over the discrete values $\omega\in\{0.5,0.75,1.0,1.25,1.5\}$, together with $h\in\{0.5,1.0,1.5\}$, for the different system sizes considered. The dependence on the driving frequency is present, although it seems to be mostly a secondary tuning effect compared to the dependence on system size. The dominant structural feature is similar to what has been shown elsewhere in the manuscript: $q=3$ is clearly the hardest scenario, while $q=2,4,5,6$ perform substantially better overall; a detailed discussion of the physical origin of this anomalous $q=3$ regime is given in~\ref{sec:appendix3}.

Regarding the 2-qubit case, the metrics generally appear to benefit from increasing $\omega$, although $\eta_{\omega}$ decreases slightly while still remaining above 0.9; this behavior is not unexpected, since this system size does not pose a significant challenge for the proposed model. On the other hand, the 3-qubit scenario with $h=1$ and $\omega=1$---standard configuration---is where the model achieves the poorest $\eta_{\omega}$ solution, whereas other parameter settings yield markedly better performance, particularly for $h=0.5$. Moreover, in this case, although $\mathcal{L}_{\mathrm{E-L}}$ remains higher than for other system sizes, it decreases as $\omega$ increases, which clearly indicates that we have encountered a particular regime that is difficult for the proposed model to resolve, though by no means impossible. For $q=4,5,6$, the performance is again generally strong, although with different degrees of robustness. In the 4-qubit case, a certain dependence on $\omega$ is still apparent, whereas for $q=5$ some choices of the $(h,\omega)$ pair are visibly less favorable. Finally, the 6-qubit case appears to be the most robust overall, with metrics remaining close to unity almost everywhere, except for a mild drop in one curve at the highest frequency value.

In general, regarding the role of $h$, there is no single universal winner for all $q$, but our results suggest that smaller $h$ is often the safest choice to maintain high QFI, especially for larger system sizes. By contrast, larger $h$ can sometimes improve fidelity for small sizes, but more often the solution is more frequency-sensitive and worsens either $\mathcal{B}$ or the loss. Therefore, the practical message is that $h$ acts as a trade-off parameter between metrological efficiency and state-targeting quality, rather than as a uniformly beneficial control. More generally, we observe that as the field becomes stronger, i.e., for larger values of $h$, the corresponding metrics become increasingly difficult to achieve. These limitations are likely to reflect not only methodological constraints, but also the intrinsic physical difficulty of the underlying dynamics.

\begin{figure}[!h]
    \centering
    \includegraphics[width=1.0\columnwidth]{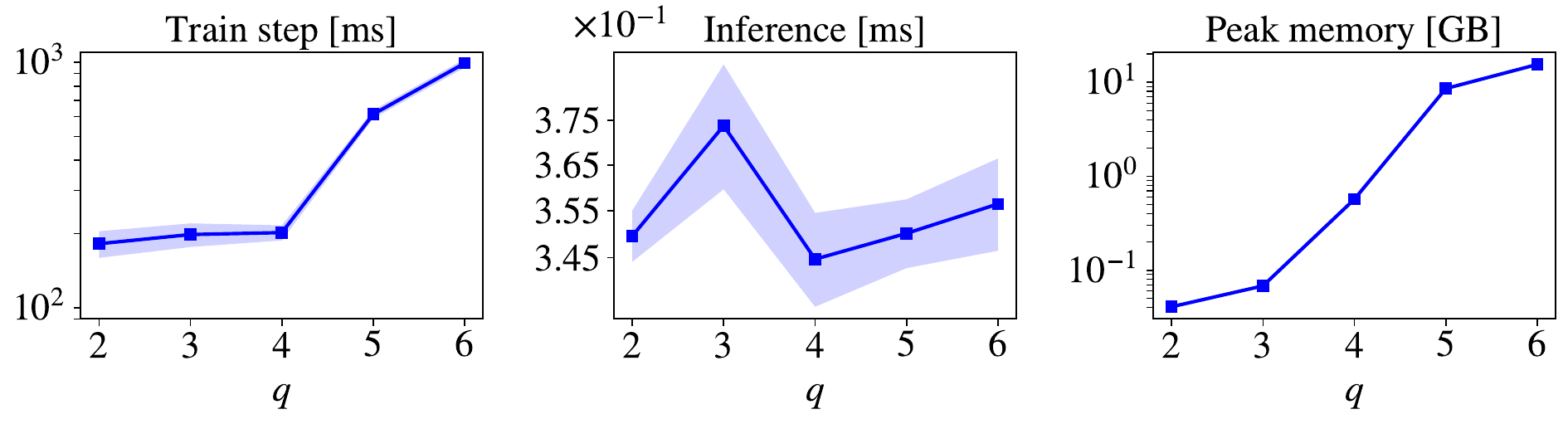}
    \caption{\justifying Training and inference computational cost of the PINN model as a function of qubit number $q$. Left: mean training time per optimization step (error bars: standard deviation across runs). Middle: mean inference time per forward pass (error bars: standard deviation across runs). Right: peak GPU memory usage during training.}
    \label{fig:scalability_1}
\end{figure}

\textbf{Scalability across system size.} When one attempts to simulate quantum dynamics on classical computers, it is natural to encounter a limitation in the system size, i.e., in the number of qubits considered. In general, the operators are represented by complex square matrices of size $2^{q}\times 2^{q}$, while the states are described by vectors of dimension $2^{q}$. Moreover, when these operators are decomposed in the basis of the corresponding projectors, as in~\eqref{eq:AGP_decomposition}, where all possible combinations of operators acting on the $q$ qubits of the system are in principle considered, their number grows up to $4^{q}$. In practice, however, when the system size becomes large, one typically restricts the description to at most two- or three-body interactions, and not even all such possibilities are retained. This helps alleviate both the computational and memory costs, since storing all matrix representations is not feasible in classical simulation. Figure~\ref{fig:scalability_1} presents the numerical study carried out, showing in the left panel the mean computational time required for one training iteration of the model, together with the $+1\sigma$ deviation, over $N=1000$ independent runs, as a function of $q$, and, similarly, the inference time in the central panel.\\

It is straightforward to see that the training time remains approximately constant for $q=2,3,4$, whereas from $q=5$ onward the expected exponential behavior becomes already apparent. It is worth noting here that for $q=6$, we consider only up to four-qubit interactions, so that the decomposition does not contain all $4^{6}=4096$ possible terms, but we should rather use the binomial coefficient as

\begin{equation}
\sum_{l=0}^{4}\frac{6!}{l!(6-l)!}3^{l}=1909
\label{eq:max_weight}
\end{equation}

\noindent allowed terms in the truncated 4-local basis\footnote{The equation in~\eqref{eq:max_weight} is fully general; by replacing the upper limit with $k$ and the numerator in the summation argument with the total number of qubits $q$, one can determine how many terms are retained in the bases, this number always being less than or equal to $4^{q}$.}. From the numbers, one can readily see that, for a standard training run of $25,000$ epochs, the model would require approximately 50 minutes on our hardware for $q=2$, whereas under our $q=6$ configuration this figure would rise to nearly 7 hours. As for the inference time, it remains at an approximately constant mean scale of $\sim 0.35$ milliseconds for all values of $q$. It is worth noting that the computational burden lies mainly in the training stage; once the model has been trained, the inference time depends primarily on the size of the network (number of parameters), as well as on the numerical precision employed. In this regard, the output dimension, namely the number of model outputs, affects the memory usage more than the inference speed itself, since the corresponding operations are vectorized. Finally, regarding the memory usage of the problems, we observe a clear exponential trend from 2 to 5 qubits, with the $q=6$ case excluded from this pattern since not all possible interactions are considered there. In this regard, these values can be qualitatively fitted by the numerical form $M(q)=ae^{bq}+c$, which yields an approximate estimate of 137 GB for the actual memory demand at $q=6$. As can be seen, this truncation reduces that expected value substantially to the practical value of $15.44$ GB reported in table~\ref{tab:scalability_summary} for all dynamics, which is already manageable.

\begin{figure}[!h]
    \centering
    \includegraphics[width=1.0\columnwidth]{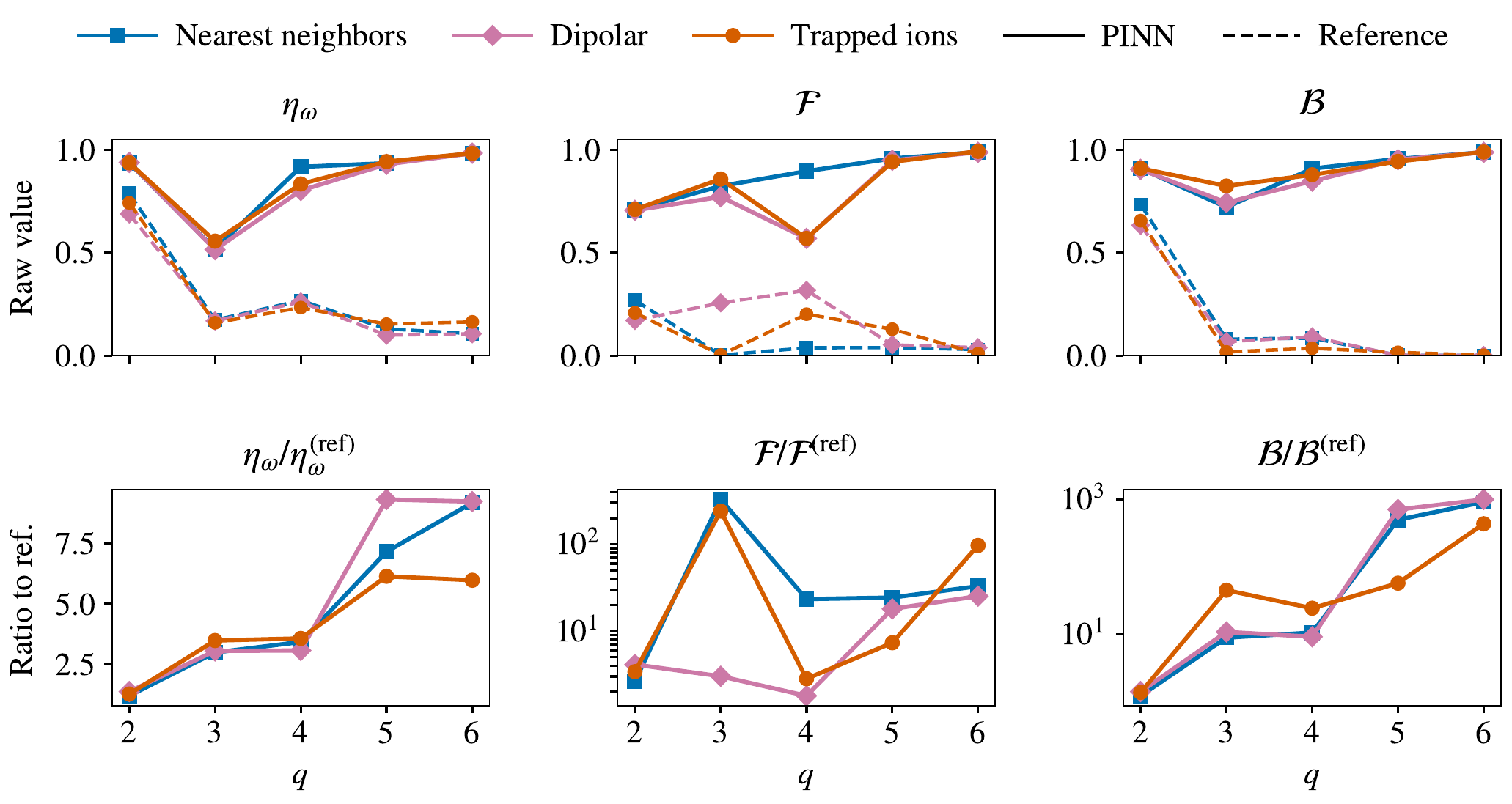}
    \caption{\justifying Physical metrics versus qubit count $q$ for the different Hamiltonians considered. The columns report $\eta_{\omega}$, $\mathcal{F}$, and $\mathcal{B}$. The top panels show the raw and reference values, while the bottom panel display the corresponding ratios with respect to that reference.}
    \label{fig:scalability_2}
\end{figure}

The bulk of the computational cost arises from the calculation of the time evolution of the quantum states, since this generally requires the use of dense matrices and their exponentiation, leading to a cost that scales as $8^{q}$. Accordingly, in assessing scalability, it is important not only to examine runtime and memory requirements, but also to quantify the performance of the methodology itself and determine whether its implementation is ultimately worthwhile. Figure~\ref{fig:scalability_2} presents the usual trio of metrics, $\eta_{\omega}$, $\mathcal{F}$, and $\mathcal{B}$, as functions of the number of qubits for the different dynamics. The panels in the first row show their raw values as solid lines, together with the corresponding reference values as dashed lines---by ``reference'' we mean those obtained when the model is trained using only $\mathcal{L}_{\mathrm{E-L}}$. The second row displays the ratio between the two, in order to make any possible gain more clearly visible.

\begin{table*}[!b]
\centering
\normalsize
\setlength{\tabcolsep}{5pt}
\renewcommand{\arraystretch}{1.12}

\caption{\textbf{Scalability summary across system size.} For the different Hamiltonians considered, and for each number of qubits $q$, we report training step time and inference time (mean $\pm$ std) across runs, peak GPU memory, and the achieved QFI efficiency $\eta_{\omega}$, final-state fidelity $\mathcal{F}$, and extremal-balance score $\mathcal{B}$.}
\label{tab:scalability_summary}

\begin{tabular*}{\textwidth}{@{\extracolsep{\fill}}@{} c |
S[table-format=4.0(2),table-align-uncertainty]
S[table-format=1.3(2),table-align-uncertainty]
S[table-format=2.2]
| S[table-format=1.3]
S[table-format=1.3]
S[table-format=1.3]
@{}}
\toprule
\(q\) &
\multicolumn{1}{c}{\textbf{Train step [ms]}} &
\multicolumn{1}{c}{\textbf{Inference [ms]}} &
\multicolumn{1}{c}{\textbf{Peak mem. [GB]}} &
\multicolumn{1}{c}{\(\eta_{\omega}\)} &
\multicolumn{1}{c}{\(\mathcal{F}\)} &
\multicolumn{1}{c}{\(\mathcal{B}\)} \\
\midrule

\multicolumn{7}{@{}l}{\textbf{Hamiltonian: Nearest neighbors}}\\
\specialrule{0.05pt}{0pt}{4pt}
2 & 128(23) & 0.349(0.005) & 0.04  & 0.938 & 0.710 & 0.910 \\
3 & 199(22) & 0.373(0.002) & 0.07  & 0.518 & 0.824 & 0.722 \\
4 & 202(14) & 0.344(0.010) & 0.57  & 0.918 & 0.897 & 0.910 \\
5 & 615(24) & 0.350(0.007) & 8.56  & 0.936 & 0.959 & 0.957 \\
6 & 994(40) & 0.356(0.010) & 15.44 & 0.984 & 0.993 & 0.990 \\
\midrule

\multicolumn{7}{@{}l}{\textbf{Hamiltonian: Dipolar}}\\
\specialrule{0.05pt}{0pt}{4pt}
2 & 185(21) & 0.451(0.003) & 0.04  & 0.940 & 0.707 & 0.906 \\
3 & 209(26) & 0.357(0.007) & 0.07  & 0.515 & 0.772 & 0.744 \\
4 & 209(13) & 0.359(0.010) & 0.57  & 0.803 & 0.570 & 0.848 \\
5 & 613(22) & 0.353(0.010) & 8.56  & 0.930 & 0.950 & 0.953 \\
6 & 997(41) & 0.346(0.008) & 15.44 & 0.984 & 0.986 & 0.989 \\
\midrule

\multicolumn{7}{@{}l}{\textbf{Hamiltonian: Trapped ions}}\\
\specialrule{0.05pt}{0pt}{4pt}
2 & 166(13) & 0.367(0.012) & 0.04  & 0.938 & 0.711 & 0.911 \\
3 & 178(15) & 0.350(0.009) & 0.07  & 0.557 & 0.859 & 0.825 \\
4 & 205(12) & 0.352(0.017) & 0.57  & 0.835 & 0.570 & 0.880 \\
5 & 617(27) & 0.359(0.015) & 8.56  & 0.944 & 0.943 & 0.944 \\
6 & 993(43) & 0.362(0.015) & 15.44 & 0.984 & 0.993 & 0.990 \\
\bottomrule
\end{tabular*}

\vspace{4pt}
\end{table*}

We find that the metrics behave largely as expected, with the recurring exceptions of $\eta_{\omega}$ for $q=3$ and $\mathcal{F}$ for $q=4$ in the dipolar and tarpped-ion dynamics. Although the raw values remain relatively stable as $q$ increases, their values relative to the reference exhibit a clear positive trend, especially in the cases of $\eta_{\omega}$ and $\mathcal{B}$. This is physically reasonable, since larger system sizes naturally allow greater room for improvement. For example, the quantity $\mathcal{B}$ reflects how strongly populated the extremal energy levels of the sensitivity operator are; as $q$ increases, the number of such levels also increases, and with it the margin of error. A similar consideration applies to the QFI.

\begin{figure}[!h]
    \centering
    \includegraphics[width=1.0\columnwidth]{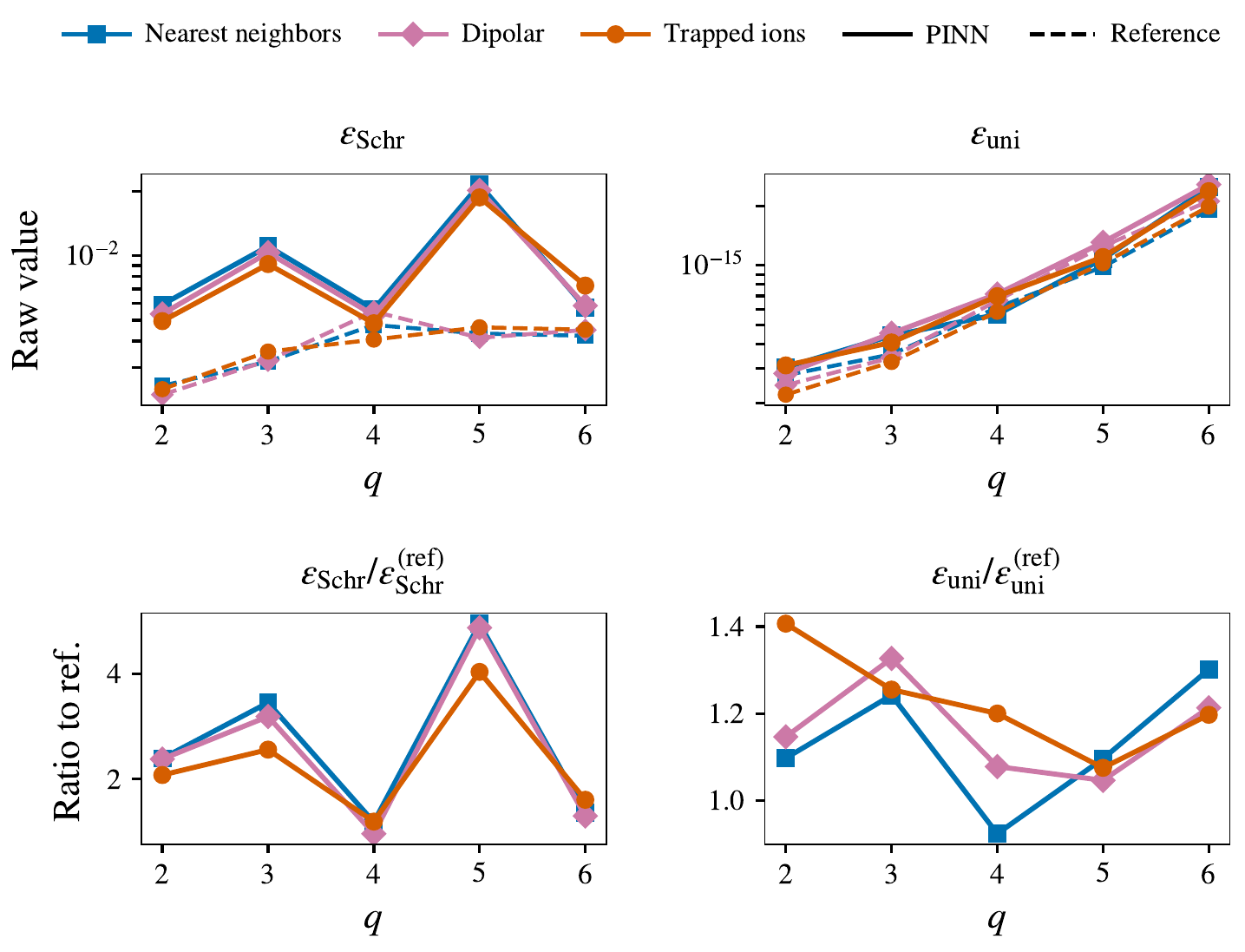}
    \caption{\justifying Scalability of the Schrödinger residual error, $\varepsilon_{\mathrm{Schr}}$, and the mean unitarity error, $\varepsilon_{\mathrm{uni}}$, as functions of the number of qubits, $q$, for the different dynamics considered. The top panels show both the raw and reference results, while the bottom panels display the corresponding rations between them.}
    \label{fig:scalability_3}
\end{figure}

On the other hand, beyond the performance of the model itself, it is also important to assess how the dynamical evolution behaves in terms of the error being made. To this end, figure~\ref{fig:scalability_3} reports the measured errors in the Schrödinger equation and in the unitarity of the time-evolution operator, as defined in~\eqref{eq:sch_metric} and~\eqref{eq:uni_error} (see Section~\ref{subsec:eval_metrics}). Once again, the first row presents the raw values of the two metrics as solid lines, while the corresponding reference values are shown as dashed lines. As can be seen, all protocols satisfy the condition that the time-averaged error $\varepsilon_{\mathrm{Schr}}$ remains below 10\%, bearing in mind that this metric is normalized with respect to the norm of the action of $\pmb{\mathcal{H}}_{\mathrm{tot}}(t)$ on the states. Consequently, if this quantity ever reached problematic values, it could always be incorporated as an additional loss term so that the model enforces it naturally. Moreover, when evaluating the ratio relative to the reference solution, we find that the largest errors occur for odd values of $q$, reflecting the greater difficulty in resolving these cases, most likely as a consequence of the symmetry properties of the particular dynamics considered. Finally, in terms of the error in the unitarity of the time-evolution operator, we find that it remains at the level of numerical precision, with errors generally of the order of $\sim 10^{-15}$. However, a clearly increasing trend with $q$ can be observed on a logarithmic scale, which seems to suggest that the temporal grid should be refined---that is, that $N_{t}$ should be increased---as the system size grows. In all numerical results reported here, we have used a fixed value of $N_{t}=2^{8}$, independently of $q$; however, this behavior appears to indicate that, for larger systems, a higher temporal resolution in the Magnus expansion is likely required. In any case, the resulting error remains entirely manageable for these scenarios.

Table~\ref{tab:scalability_summary} reports the numerical results for all these calculations, where the trend in the maximum memory usage can be clearly appreciated. This quantity is the same regardless of the dynamics considered, as expected, since all of them are equally costly to construct. Up to minor variations, the same trend is also observed in the training time per iteration, though not in the inference time, which remains approximately constant through the tests. Nevertheless, the final output dimension of the PINN scale as~\eqref{eq:size_output}. Taking into account the number of outputs, it is possible to express exactly, mathematically, the amount of memory they occupy in GiB, according to~\eqref{eq:memory_occupation}, where $s_{\mathrm{dtype}}$ is a parameter determined by the numerical precision employed. In our case, since all outputs are real-valued, this corresponds to float32, that is, $s_{\mathrm{dtype}}=s_{\mathrm{float32}}=4$ bytes.

\begin{equation}
M_{\mathrm{out}}^{\mathrm{GiB}}(q,k)=\frac{N_{t}}{1024^{3}}N_{\mathrm{out}}(q,k)\:s_{\mathrm{dtype}}.
\label{eq:memory_occupation}
\end{equation}

\begin{figure}[!h]
    \centering
    \includegraphics[width=1.0\columnwidth]{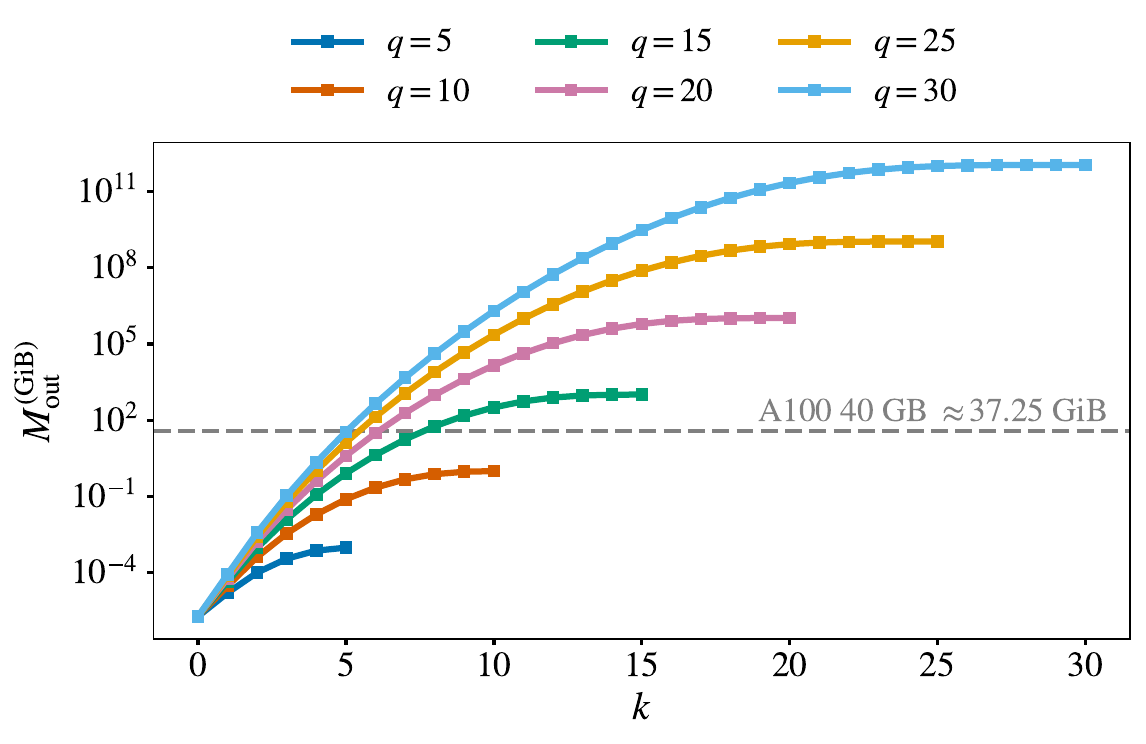}
    \caption{\justifying Memory occupation of the PINN output tensor, $M_{\mathrm{out}}^{\mathrm{GiB}}$, as a function of the locality cutoff $k$, for different system sizes $q$. The curves are computed from the exact output dimension $N_{\mathrm{out}}(q,k)$, assuming float32 precision and $N_{t}=2^{8}$ collocation points.}
    \label{fig:memory_output}
\end{figure}

It is crucial to understand, however, that this term alone does not account for the total memory used by the model. One must also include the memory associated with the trainable parameters themselves, $M_{\mathcal{NN}}(q,k)$, as well as the memory required by PyTorch automatic differentiation, $M_{\mathrm{autograd}}(q,k)$, which in turn depends on factors such as the optimizer employed and the precision of the complex tensors themselves, represented in complex128 precision. These additional contributions are considerably more difficult to estimate accurately, and for this reason we restrict our analysis to the memory associated solely with the existence of the outputs, without considering any further components. This quantity should therefore be interpreted as a lower bound on the required memory, since aspects such as network size, optimizer, or differentiation strategy may always be adapted to reduce computational overhead, while the number of outputs is an intrinsic feature that cannot be altered in any way except by the cutoff $k$. In figure~\ref{fig:memory_output}, we draw $M_{\mathrm{out}}^{\mathrm{GiB}}(q,k)$ as a function of $k$ for different values of $q$, together with the GiB limit corresponding to the 40 GB A100 GPU card used in our experiments. As can be seen, if one were to store exclusively the AGP coefficients, a clear limit would already emerge at a locality of approximately $k=5\sim 6$ for system sizes $q\geq 15$. In practice, as noted above, this threshold would be substantially lower, since the bulk of the computational overhead arises from the automatic differentiation required to compute the gradient of the loss function with respect to all trainable parameters. Even so, it is instructive to observe that this would constitute the practical limitation in the case where only the AGP itself is represented---we ignore here $\lambda(t)$ as output since its contribution would be negligible in this scenario---, this being the bare minimum required.

\section{Conclusions}
\label{sec:Conclusions}

From a physical standpoint, the problem addressed in this work is the design of time-dependent quantum dynamics capable of encoding a parameter with the highest possible sensitivity, as quantified by the quantum Fisher information (QFI). This is a central objective in quantum metrology, since the QFI sets the ultimate precision bound for parameter estimation and therefore determines how efficiently a quantum system can be used as a sensor. In interacting many-body settings, however, this optimization problem becomes highly non-trivial because the dynamics, the spectral structure, and the control requirements are all strongly constrained by non-commutativity and by the exponential growth of the Hilbert space. In this context, we introduce a physics-informed neural network (PINN) framework to learn counter-diabatic (CD) dynamics in time-dependent many-body systems. Within this approach, the adiabatic gauge potential (AGP) and the associated control protocol are inferred in a way that enhances the dynamical generator underlying the quantum Fisher information (QFI), thereby increasing its variance and improving parameter sensitivity. In this way, the adiabatic gauge potential (AGP), written as $\pmb{\mathcal{A}}_{\lambda}(t)$, and the associated control protocol are inferred within a metrologically driven but physically consistent setting.

The numerical results show that the proposed framework systematically outperforms reference solutions based solely on Euler–Lagrange optimization, highlighting the importance of explicitly incorporating metrological objectives, rather than purely dynamical criteria, into the control design.
In most of the cases considered, the model reaches high values of the normalized QFI ($\eta_{g}$) together with favorable fidelity ($\mathcal{F}$) and the introduced extremal-balance score ($\mathcal{B}$), while maintaining small physical residuals. These results indicate that the explicit incorporation of metrological objectives into the loss function is essential, since minimizing the physical action alone is generally insufficient to maximize the QFI. This is because it does not directly control the variance of the generator associated with parameter encoding, which ultimately determines the achievable precision.
The analysis also highlights the relevance of learning the scheduling function rather than fixing it a priori. In general, allowing $\lambda(t)$ to be trainable provides an additional degree of freedom that improves the final performance and, in many cases, also yields a better compromise between metrological gain and physical consistency. At the same time, this benefit is not entirely uniform across all Hamiltonians and system sizes, which points to a non-trivial dependence on the spectral and dynamical structure of the underlying problem.

A further relevant observation is that the performance does not vary monotonically with the size of the system. In particular, the $q=3$ case appears consistently as one of the most challenging scenarios, with a milder degradation also present in some $q=4$ cases. As discussed in~\ref{sec:appendix3}, this behavior is not explained solely by computational overhead but is instead consistent with a finite-size mismatch between the symmetry structure of the metrologically relevant subspace and that of the available dynamics.

From the computational perspective, the Magnus expansion constitutes a central ingredient of the framework, since it makes the treatment of time-ordered dynamics during training considerably more tractable while preserving a good level of accuracy. Nevertheless, scalability remains the main limitation of the method. Even when the gauge potential is represented symbolically through Pauli-string coefficients and locality restrictions are imposed, the size of the operator space and the associated memory demands still grow very rapidly with the number of qubits, in particular once automatic differentiation is taken into account. It should also be noted that the present study has been restricted to the specific Hamiltonian families considered here, namely the chosen nearest-neighbor, dipolar, and trapped-ion-inspired dynamics within the corresponding driving structure. Although these examples provide a meaningful validation of the methodology in non-trivial many-body settings, they do not consider the broader landscape of dynamical scenarios relevant to the quantum metrology field. For this reason, an important direction for future work will be to extend the analysis to different Hamiltonian classes, broader dynamical regimes, and alternative control paradigms, in order to assess whether the good performance observed in this work persists across qualitatively different physical settings. In particular, it would be of clear interest to determine whether the model remains equally effective when applied to different interaction patterns, different control landscapes, or more general metrological protocols. Beyond the system sizes studied here, the proposed framework can be viewed as a hybrid classical-quantum design tool for learning physically structured control ansätze, reduced AGP models, and scheduling strategies that may later be transferred to larger-scale simulations or hardware-oriented implementations. Thus, the present $2-6$-qubit results are not only a proof of concept, but also a phyiscally grounded step toward scalable control design for quantum metrology.

In summary, the results support the view that PINNs constitute a viable and physically grounded framework for metrological quantum control in interacting systems. While significant challenges remain, particularly in relation to scalability and generalization across problem classes, the methodology shows clear potential as a flexible route for learning time-dependent control strategies capable of approaching the fundamental limits of precision.

\section*{Acknowledgments}
\label{sec:Acknowledgments}
The authors gratefully acknowledge the computer resources at Artemisa, funded by the European Union ERDF and Comunitat Valenciana as well as the technical support provided by the Instituto de Fisica Corpuscular, IFIC (CSIC-UV). This work has been partially supported by the Valencian Government Grant No. CIAICO/2024/111 and the Spanish Ministry of Economic Affairs and Digital Transformation through the QUANTUM ENIA project call – Quantum Spain project, and the European Union through the Recovery, Transformation and Resilience Plan – NextGenerationEU (Digital Spain 2026) Agenda. X.C. and Y.B. appreciate 
PID2024-157842OA-I00 and PID2021-126273NB-I00 funded by MCIN/AEI/10.13039/501100011033 and by ``ERDF A way of making Europe'' and ``ERDF Invest in your Future'', the Basque Government through Grant No. IT1470-22, and the Severo Ochoa Centres of Excellence program through Grant CEX2024-001445-S. Y.B. also acknowledges 
the project in the field of Artificial Intelligence 2025 (AIA2025-163435-C44).

\section*{Data and code availability statement}
\label{sec:data_and_code}

The code and data used to reproduce the results reported in this work, together with the trained models, will be made available by the authors upon reasonable request from researchers until publication of the article in a journal. After publication, these materials will be deposited in a public repository and made openly available.


\printbibliography

\appendix
\renewcommand{\thesection}{Appendix~\Alph{section}}
\section{Magnus expansion}
\label{sec:appendix1}
\counterwithin*{equation}{section}
\renewcommand{\theequation}{A.\arabic{equation}}

The Magnus expansion provides a convenient framework for computing the time evolution of a quantum state from one instant to another by expressing the time-evolution operator as the exponential of an effective operator, denoted by $\pmb{\Omega}$. This operator admits a series representation, where the resulting approximation order is determined by the truncation order $p$ considered. After partitioning the time axis into $n_{w}$ windows, an operator $\pmb{\Omega}^{(w)}$ can be computed for each window, which propagates the state from the beginning to the end of that window, as expressed in~\eqref{eq:Magnus_time_evolution}. Taking $\pmb{\mathcal{H}}_{\mathrm{tot}}(t)$ as the Hamiltonian governing the dynamics, the first three discrete orders of $\pmb{\Omega}^{(w)}$ are reported below.

\begin{equation}
\begin{aligned}
\pmb{\Omega}_{1}^{(w)} \;&=\; -i\,\Delta t \sum_{j=0}^{m-1} \pmb{\mathcal{H}}_{\mathrm{tot}}\!\left(t^{(w)}_{j}\right),\\[6pt]
\pmb{\Omega}_{2}^{(w)} \;&=\; -\frac{\Delta t^{2}}{2}\sum_{j_{1}=0}^{m-1}\sum_{j_{2}=0}^{j_{1}-1}
\Bigl[\pmb{\mathcal{H}}_{\mathrm{tot}}\!\left(t^{(w)}_{j_{1}}\right),\,\pmb{\mathcal{H}}_{\mathrm{tot}}\!\left(t^{(w)}_{j_{2}}\right)\Bigr],\\[6pt]
\pmb{\Omega}_{3}^{(w)} \;&=\; \frac{i\,\Delta t^{3}}{6}\sum_{j_{1}=0}^{m-1}\sum_{j_{2}=0}^{j_{1}-1}\sum_{j_{3}=0}^{j_{2}-1}
\Bigl(
\Bigl[\pmb{\mathcal{H}}_{\mathrm{tot}}\!\left(t^{(w)}_{j_{1}}\right),\Bigl[\pmb{\mathcal{H}}_{\mathrm{tot}}\!\left(t^{(w)}_{j_{2}}\right),\pmb{\mathcal{H}}_{\mathrm{tot}}\!\left(t^{(w)}_{j_{3}}\right)\Bigr]\Bigr]\\
&\hspace{2.6cm}+
\Bigl[\pmb{\mathcal{H}}_{\mathrm{tot}}\!\left(t^{(w)}_{j_{3}}\right),\Bigl[\pmb{\mathcal{H}}_{\mathrm{tot}}\!\left(t^{(w)}_{j_{2}}\right),\pmb{\mathcal{H}}_{\mathrm{tot}}\!\left(t^{(w)}_{j_{1}}\right)\Bigr]\Bigr]
\Bigr).
\end{aligned}
\label{eq:Magnus_terms}
\end{equation}

The error in the time evolution caused by truncating the expansion can be derived as follows. Let $\pmb{\mathcal{U}}(T)=\prod_{w=1}^{n_{w}}\pmb{\mathcal{U}}^{(w)}$ be the exact evolution split into $n_{w}$ windows of width $h=T/n_{w}$, where each exact window propagator is $\pmb{\mathcal{U}}^{(w)}=\mathcal{T}\:\mathrm{exp}\left(-i\int_{t^{(w-1)}}^{t^{(w)}}\pmb{\mathcal{H}}_{\mathrm{tot}}(t)dt\right)$. The Magnus expansion writes $\pmb{\mathcal{U}}^{(w)}=\mathrm{exp}\left(\pmb{\Omega}^{(w)}\right)$ with $\pmb{\Omega}^{(w)}=\sum_{n\geq 1}\pmb{\Omega}_{n}^{(w)}$, and a $p$-th truncation order uses $\pmb{\tilde{\mathcal{U}}}^{(w)}=\mathrm{exp}(\pmb{\tilde{\Omega}}^{(w)})$ where $\pmb{\tilde{\Omega}}^{(w)}=\sum_{n=1}^{p}\pmb{\tilde{\Omega}}_{n}^{(w)}$. The local (per-window) defect is governed by the first neglected term, that is,

\begin{equation}
\pmb{\Omega}^{(w)}-\pmb{\tilde{\Omega}}^{(w)}=\sum_{n\geq p+1}\pmb{\Omega}_{n}^{(w)}.
\end{equation}

Standard Magnus estimates give $||\pmb{\Omega}_{n}^{(w)}||\leq C_{n}h^{n}$, where $C_{n}$ depends on supremum norms of $-i\pmb{\mathcal{H}}_{\mathrm{tot}}$ and its $(n-1)$-fold nested commutators on the window. Hence,

\begin{equation}
||\pmb{\Omega}^{(w)}-\pmb{\tilde{\Omega}}^{(w)}||\leq C_{p+1}h^{p+1}+\mathcal{O}\left(h^{p+2}\right).
\end{equation}

Using $||e^{X}-e^{Y}||\leq e^{\mathrm{max}\left(||X||,||Y||\right)}||X-Y||$, we get a per-window propagator error $||\pmb{\mathcal{U}}^{(w)}-\pmb{\tilde{\mathcal{U}}}^{(w)}||\leq K^{(w)}C_{p+1}h^{p+1}+\mathcal{O}\left(h^{p+2}\right)$ for some bounded $K^{(w)}$. Propagating across windows (a product argument) yields a global final-time error scaling as

\begin{equation}
||\pmb{\mathcal{U}}(T)-\pmb{\tilde{\mathcal{U}}}(T)||\lesssim\sum_{w=1}^{n_{w}}||\pmb{\mathcal{U}}^{(w)}-\pmb{\tilde{\mathcal{U}}}^{(w)}||\leq Kn_{w}\:h^{p+1}=KTh^{p}=\mathcal{O}\left(\frac{T^{p+1}}{n_{w}^{p}}\right),
\label{eq:theo_error_Magnus_expansion}
\end{equation}

\noindent with $K$ collecting commutator-norm constants and stability factors. If each window contains $m$ fine time steps of size $\Delta t$ (so $h=m\Delta t$ and $n_{w}=T/(m\Delta t)$, the same scaling becomes

\begin{equation}
||\pmb{\mathcal{U}}(T)-\pmb{\tilde{\mathcal{U}}}(T)||=\mathcal{O}\left(T(m\Delta t)^{p}\right),
\end{equation}

\noindent i.e., higher Magnus order $p$ improves the power of the window size, and using fewer/larger windows (larger $h$) increases the error accordingly.

It is important to highlight that the Magnus propagator adopted here is only one possible choice, namely a commutator-based truncation of the exponential generator. Other alternatives are also available,such as commutator-free exponential time propagators, which approximate the evolution as products of exponentials of weighted Hamiltonians without explicitly evaluating nested commutators~\cite{Alvermann_2011}, or split-operator schemes when the operator admits a suitable decomposition~\cite{Feit_1982}. The most convenient choice is therefore problem-dependent.

\section{Euler-Lagrange residual in Pauli-coefficient form}
\label{sec:appendix2}
\counterwithin*{equation}{section}
\renewcommand{\theequation}{B.\arabic{equation}}

In the numerical implementation, the Euler-Lagrange residual is not evaluated by constructing the full operators as dense matrices, but directly in the Pauli-string basis. Let $\{P_{k}\}_{k=1}^{M}$ denote the base of the chosen operator. Then the adiabatic gauge potential and the control Hamiltonian are expanded as

\begin{equation}
\pmb{\mathcal{A}}_{\lambda}(t)=\sum_{i=1}^{M}a_{i}(t)P_{i},\quad\pmb{\mathcal{H}}_{g}(t)=\sum_{j=1}^{M}h_{j}(t)P_{j},\quad\partial_{\lambda}\pmb{\mathcal{H}}_{g}(t)=\sum_{i=1}^{M}g_{i}(t)P_{i}.
\end{equation}

The Euler-Lagrange operator entering the loss is

\begin{equation}
\mathrm{EL}(t)\equiv\left[i\partial_{\lambda}\pmb{\mathcal{H}}_{g}(t)-\left[\pmb{\mathcal{A}}_{\lambda}(t),\pmb{\mathcal{H}}_{g}(t)\right],\pmb{\mathcal{H}}_{g}(t)\right].
\end{equation}

The key point is that commutators can also be written in the same basis. For two basis elements $P_{i}$ and $P_{j}$, one has

\begin{equation}
[P_{i},P_{j}]=\sum_{k=1}^{M}C_{ij}^{k}P_{k},
\end{equation}

\noindent where $C_{ij}^{k}$ are the structure constants of the chosen Pauli basis. Therefore,

\begin{equation}
\left[\pmb{\mathcal{A}}_{\lambda}(t),\pmb{\mathcal{H}}_{g}(t)\right]=\sum_{i,j=1}^{M}a_{i}(t)\:h_{j}(t)\:[P_{i},P_{j}]=\sum_{k=1}^{M}\left(\sum_{i,j=1}^{M}C_{ij}^{k}\:a_{i}(t)\:h_{j}(t)\right)P_{k}.
\end{equation}

Hence, the coefficients of the first commutator can be identified as

\begin{equation}
c_{k}(t)=\sum_{i,j}^{k}C_{ij}^{k}\:a_{i}(t)\:h_{j}(t).
\end{equation}

Defining $q_{k}(t)\equiv ig_{k}(t)-c_{k}(t)$, the intermediate operator becomes

\begin{equation}
i\partial_{\lambda}\pmb{\mathcal{H}}_{g}(t)-\left[\pmb{\mathcal{A}}_{\lambda}(t),\pmb{\mathcal{H}}_{g}(t)\right]=\sum_{k=1}^{M}q_{k}(t)\:P_{k}.
\end{equation}

Applying the second commutator with $\pmb{\mathcal{H}}_{g}(t)$, we obtain

\begin{equation}
\mathrm{EL}(t)=\sum_{k=1}^{M}r_{k}(t)\:P_{k},\quad r_{k}(t)=\sum_{i,j=1}^{M}C_{ij}^{k}\:q_{i}(t)\:h_{j}(t).
\end{equation}

Equivalently, after substituting $q_{i}(t)$,

\begin{equation}
r_{k}(t)=\sum_{i,j=1}^{M}C_{ij}^{k}\left(i\:g_{i}(t)-c_{i}(t)\right)\:h_{j}(t).
\end{equation}

These are precisely the coefficients computed numerically in the implementation. Here, $M$ denotes the number of Pauli strings retained in the operator basis used in the numerical implementation, i.e., $\{P_{1},\ldots,P_{M}\}$. Since the full operator space of a $q$-qubit system contains $4^{q}$ Pauli strings, one works with a truncated basis and projects the commutators back onto that space whenever terms outside the retained set are generated. This truncation makes the method tractable numerically without reducing its generalization power, since the model is still trained to satisfy the Euler-Lagrange equation over the whole time domain within a physically motivated operator subspace, which captures the relevant structure of the problem rather than a single instance of the dynamics.

\section{On the anomalous $q=3$ regime}
\label{sec:appendix3}
\counterwithin*{equation}{section}
\renewcommand{\theequation}{C.\arabic{equation}}

A recurrent feature of the results in table~\ref{tab:results_summary} is that the $q=3$ case systematically achieves lower $\eta_{\omega}$, and in general also less performance in terms of $\mathcal{F}$ and $\mathcal{B}$, than the neighboring system sizes. Since the metrological upper bound is determined by the extremal eigenspaces of the sensitivity operator $\partial_{\omega}\pmb{\mathcal{H}}_{\omega}(t)$, and the ideal protocol would keep the evolving state aligned with the balanced superposition of those instantaneous extremal states, it is natural to examine whether the learned dynamics is less capable of tracking that subspace precisely in this finite-size regime. To this end, we define the projector onto the instantaneous extremal subspace,

\begin{equation}
\pmb{\Pi}_{\mathrm{ext}}(t)=\ket{\phi_{\mathrm{min}}(t)}\bra{\phi_{\mathrm{min}}(t)}+\ket{\phi_{\mathrm{max}}(t)}\bra{\phi_{\mathrm{max}}(t)},
\label{eq:ext_proj}
\end{equation}

and the corresponding extremal-subspace population of the learned state,

\begin{equation}
P_{\mathrm{ext}}(t)=\bra{\Psi(t)}\pmb{\Pi}_{\mathrm{ext}}(t)\ket{\Psi(t)}=\left|\braket{\Psi(t)|\phi_{\mathrm{min}}(t)}\right|^{2}+\left|\braket{\Psi(t)|\phi_{\mathrm{max}}(t)}\right|^{2}.
\label{eq:P_ext}
\end{equation}

\begin{figure}[!h]
    \centering
    \includegraphics[width=1.0\columnwidth]{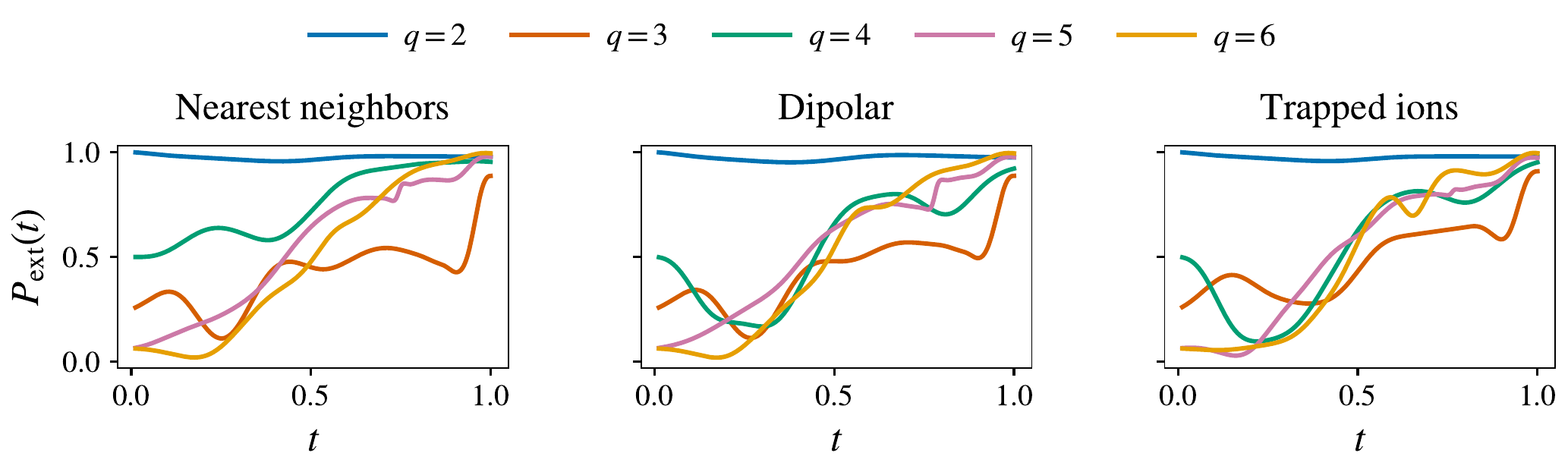}
    \caption{\justifying Time-resolved extremal-subspace population $P_{\mathrm{ext}}(t)$ for the three Hamiltonian families and system sizes $q=2,\ldots,6$. This quantity measures the total population of the learned state inside the instantaneous extremal eigenspace of the sensitivity operator.}
    \label{fig:P_ext}
\end{figure}

This quantity measures how much of the evolved state remains inside the two dimensional subspace that is metrologically relevant at each time step. Figure~\ref{fig:P_ext} shows that, although the learned dynamics does not remain fully confined to that subspace for any nontrivial system size, the $q=3$ case achieves a systematically poorer overlap across the three Hamiltonian scenarios considered. In particular, this overlap remains substantially lower over larger portions of the evolution that for $q=4,5,6$, indicating that the 3-qubit regime has a greater tendency to leak into non-optimal directions of the Hilbert space.

This observation should be interpreted comparatively rather than as a strict pointwise optimality condition. Condition in~\eqref{eq:Psi_def} represents an idealized protocol saturating the metrological bound, but in the present PINN framework neither the loss function nor the numerical propagation enforces exact confinement to the instantaneous extremal subspace at all intermediate times. From a practical metrological point of view, what really matters is that the learned dynamics reaches a sufficiently favorable final state while remaining physically consistent, not that it exactly follows the instantaneous extremal manifold at every time step. For this reason, the role of the $P_{\mathrm{ext}}(t)$ metric is diagnostic: it reveals that the 3-qubit system tracks the metrologically relevant subspace worse than the other sizes, even though perfect tracking is not achieved either for bigger systems.

To explore whether this anomalous behavior has a physical origin beyond optimization alone, we further examine the mismatch with respect to the global operator

\begin{equation}
\pmb{S}_{\mathrm{X}}=\bigotimes_{j=1}^{q}\sigma_{\mathrm{X}}^{(j)},
\label{eq:S_x}
\end{equation}

through the normalized commutator magnitude

\begin{equation}
\widetilde{\mathcal{C}}_{\mathcal{O}}(t;\pmb{S}_{\mathrm{X}})=\frac{||[\pmb{\mathcal{O}}(t),\pmb{S}_{\mathrm{X}}]||_{F}}{||\pmb{\mathcal{O}}(t)||_{F}||\pmb{S}_{\mathrm{X}}||_{F}}.
\label{eq:tilde_C}
\end{equation}

\begin{figure}[!h]
    \centering
    \includegraphics[width=1.0\columnwidth]{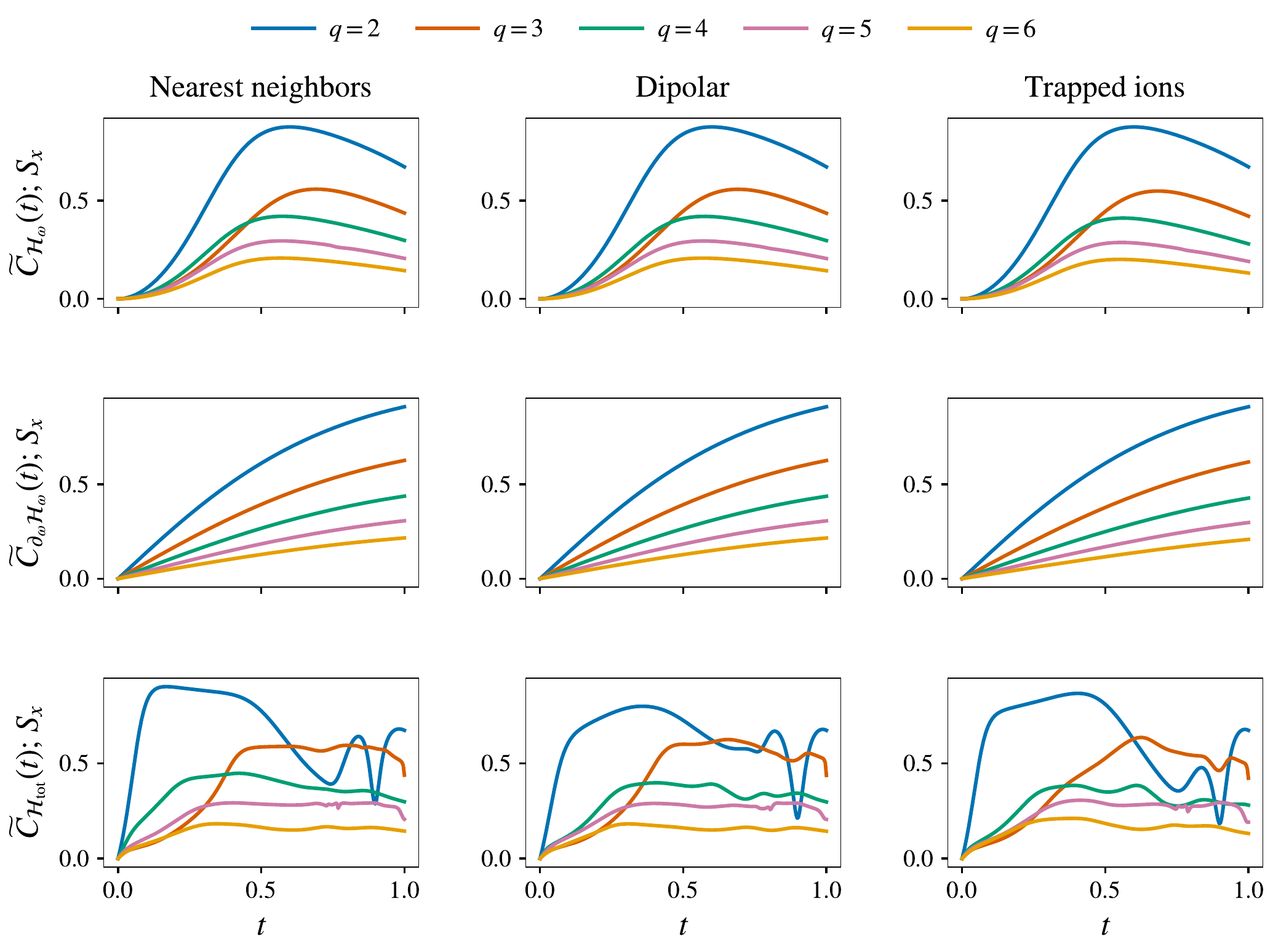}
    \caption{\justifying Normalized $\pmb{S}_{\mathrm{X}}$-symmetry mismatch $\widetilde{\mathcal{C}}_{\mathcal{O}}(t;\pmb{S}_{\mathrm{X}})$ for the different operators considered, across the three Hamiltonian families and system sizes $q=2,\ldots,6$. Among the nontrivial sizes $q\geq 3$, the $q=3$ scenario consistently gives the strongest mismatch, suggesting a physical incompatibility between the symmetry structure of the metrologically relevant subspace and that of the available dynamics.}
    \label{fig:simmetries_Sx}
\end{figure}

Figure~\ref{fig:simmetries_Sx} reports this quantity for $\pmb{\mathcal{O}}(t)=\pmb{\mathcal{H}}_{\omega}(t)$, $\partial_{\omega}\pmb{\mathcal{H}}_{\omega}(t)$, and $\pmb{\mathcal{H}}_{\mathrm{tot}}(t)$. The results show that, among the sizes $q\geq 3$, the $q=3$ case shows the largest mismatch with respect to $\pmb{S}_{\mathrm{X}}$, consistently across the three Hamiltonian families. This is particularly relevant because the effect is not restricted to the bare Hamiltonian $\pmb{\mathcal{H}}_{\omega}(t)$, but is also present in the sensitivity operator $\partial_{\omega}\pmb{\mathcal{H}}_{\omega}(t)$, being this the operator that determines the extremal states entering the metrological construction, and remains comparatively high for the learned $\pmb{\mathcal{H}}_{\mathrm{tot}}(t)$.

Taken together, these results suggest that the anomalous $q=3$ regime is not explained simply by a smaller extremal gap, but rather by a stronger incompatibility between the symmetry structure of the metrologically relevant subspace and that of the available dynamics. In that sense, the $q=3$ case appears to be a particularly unfavorable finite-size configuration in which the learned protocol must reconstruct the optimal extremal superposition under conditions of enhanced $\pmb{S}_{\mathrm{X}}$-symmetry mismatch, making the tracking problem more restrictive than in the other cases.

Finally, the $q=2$ case must be regarded as a special low-dimensional scenario. Although its normalized commutator magnitudes are in some cases larger than those of the many-body configurations, this does not translate into a harder control problem because the Hilbert space is minimal, the number of leakage channels is strongly reduced, and the learned protocol can effectively cover the relevant dynamical manifold. Therefore, $q=2$ is easier to solve despite its comparatively large apparent symmetry mismatch, whereas $q=3$ is the first genuinely nontrivial size for which that mismatch becomes dynamically harmful.

\end{document}